\documentclass[preprint,aps,prd,floatfix,superscriptaddress,preprintnumbers,nofootinbib]{revtex4}

\pdfoutput=1

\usepackage{graphicx}
\usepackage{amssymb}
\usepackage{amsmath,amssymb}
\usepackage{color}
\usepackage{multirow}

\newcommand{\beq}{\begin{equation}}
\newcommand{\eeq}{\end{equation}} 

\usepackage[bookmarks=false]{hyperref}

\allowdisplaybreaks

\begin{document}

\preprint{FERMILAB-PUB-13-319-T}

\title{Low Energy Probes of PeV Scale Sfermions }

\author{Wolfgang Altmannshofer}
\affiliation{Fermi National Accelerator Laboratory, P.O. Box 500, Batavia, IL 60510, USA}
\author{Roni Harnik}
\affiliation{Fermi National Accelerator Laboratory, P.O. Box 500, Batavia, IL 60510, USA}
\author{Jure Zupan}
\def\Cincy{Department of Physics, University of Cincinnati, Cincinnati, Ohio 45221,USA}
\affiliation{\Cincy}

\begin{abstract}
We derive bounds on squark and slepton masses in mini-split supersymmetry scenario using low energy experiments. In this setup gauginos are at the TeV scale, while sfermions are heavier by a loop factor. We cover the most sensitive low energy probes including electric dipole moments~(EDMs), meson oscillations and charged lepton flavor violation (LFV) transitions. A leading log resummation of the large logs of gluino to sfermion mass ratio is performed. A sensitivity to PeV squark masses is obtained at present from kaon mixing measurements.  A number of observables, including neutron EDMs, $\mu$ to $e$ transitions and charmed meson mixing, will start probing sfermion masses in the 100~TeV-1000~TeV range with the projected improvements in the experimental sensitivities. We also discuss the implications of our results for a variety of models that address the flavor hierarchy of quarks and leptons. We find that EDM searches will be a robust probe of models in which fermion masses are generated radiatively, while 
LFV searches remain sensitive to simple-texture based flavor models.   
\end{abstract}

\maketitle

\section{Introduction}
The LHC has placed natural models of supersymmetry (SUSY) under some stress. Direct limits for gluinos are now beyond a TeV and even direct searches for stops have improved to above 500~GeV in many scenarios. Furthermore, the discovery of a Higgs with a mass of $\sim$126 GeV is difficult to accommodate within the minimal supersymmetric standard model (MSSM) with a natural spectrum. In the MSSM the tree level Higgs mass is below the $Z$ mass and loops of top squarks are required to push the Higgs mass to the observed value. This contribution scales only logarithmically with the stop mass and as a result the top superpartner is now compelled to be heavy. This is in stark contrast with what is required to tame the quadratic divergence in the Higgs potential for which a light stop is needed. Though it is premature to exclude a natural supersymmetric spectrum, this situation has drawn attention to 
the possibility that perhaps SUSY does not address the hierarchy problem fully, but merely ameliorates it significantly. An attractive scenario along these lines is one in which SUSY is slightly split, with gauge superpartners around a TeV, and only fully 
restored around 100-1000 TeV~\cite{Giudice:1998xp, Wells:2004di, Ibe:2006de, Acharya:2007rc, Acharya:2008zi, Hall:2011jd, Ibe:2011aa, Ibe:2012hu, Acharya:2012tw, Bhattacherjee:2012ed, Arvanitaki:2012ps, ArkaniHamed:2012gw}. 
If this is the case, the Higgs mass can be accommodated easily and the fine tuning of the EW scale is a mere one part in $10^4-10^6$, a clear improvement on the 30 orders of magnitude hierarchy problem in the standard model (SM).

A PeV splitting of SUSY, a.k.a. mini-split SUSY, has several notable qualities. Before the Higgs discovery, split SUSY~\cite{ArkaniHamed:2004fb,Giudice:2004tc,ArkaniHamed:2004yi} could live in a very wide parameter space with scalar masses possibly near the GUT scale. Achieving large splittings between scalars and gauginos required, however, elaborate model building~\cite{ArkaniHamed:2004fb}. The problem is that anomaly mediation~\cite{Giudice:1998xp} gives gauginos a mass which is only a loop factor below the mass of the gravitino and is difficult to ``turn off''. Now that we know the Higgs mass is $\sim$126~GeV, such a widely split SUSY is disfavored. Instead, 100-1000~TeV is enough and split SUSY can live where it is happiest -- with SUSY breaking mediated to scalars by Planck suppressed operators and with anomaly mediation~\cite{Giudice:1998xp}\footnote{This is the case in which scalar masses are non-sequestered as opposed to the sequestered kind~\cite{Randall:1998uk}.} giving gauginos mass a loop factor 
below. 

\begin{figure}[tb] \centering
\includegraphics[width=0.96\textwidth]{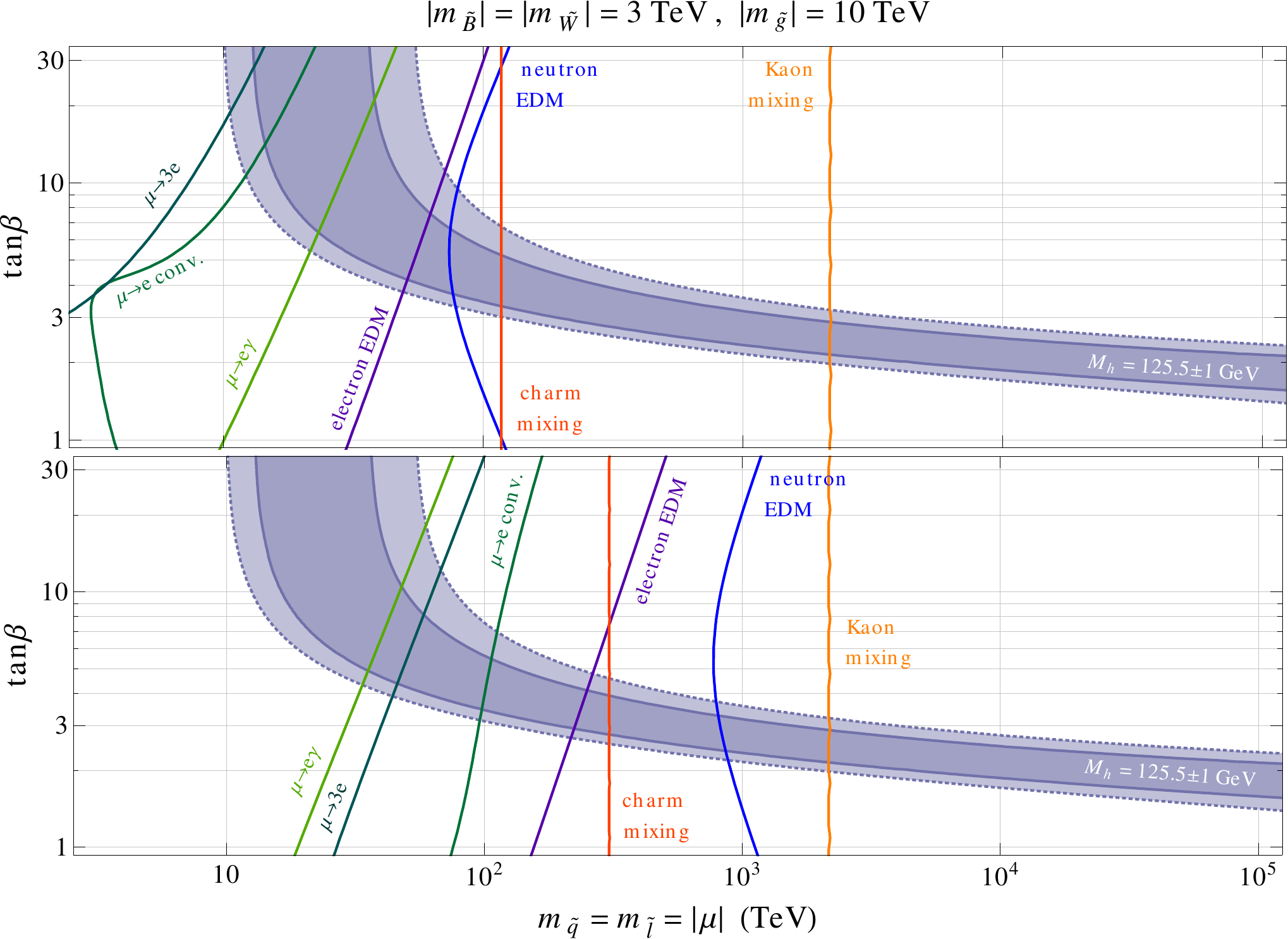}
\caption{\small
Summary of various low energy constraints (left of the lines are the excluded regions) in the sfermion mass vs. $\tan\beta$ plane for the example of 3 TeV bino and wino and 10 TeV gluino, while fixing the mass insertion parameters to be $(\delta_A)_{ij}=0.3$ when using the super-CKM basis. The dark (light) blue shaded band is the parameter space compatible with a Higgs mass of $m_h = 125.5\pm 1$~GeV within 1$\sigma$ ($2\sigma$). 
The upper (lower) plot gives the reach of current (projected future) experimental results collected in Tab.~\ref{tab:experiments}.
}
\label{fig:Mh}
\end{figure}

SUSY breaking which is mediated by Planck suppressed operators is well known to generically violate flavor in the squark and slepton sectors. Interestingly, the 100-1000~TeV scale is being probed by current and upcoming searches for flavor violating processes  and searches for electric dipole moments (EDMs). In this work we investigate the limits that these searches place on flavor violation at the PeV scale. We will see that in many cases the diagrams which constrain the split SUSY case are different than those which place constraints in the well studied low scale SUSY case. Our results are summarized in Fig.~\ref{fig:Mh} in which current bounds and future sensitivity to the scalar masses is shown in a slice of parameter space (see the next section for more details of assumptions made).  
Our conclusion is that the 0.1-1 PeV scale will be probed by a host of experiments in the near future. Constraints from Kaon oscillations are already probing squark masses of a PeV. Bounds on neutron and nuclear EDMs are likely to improve by several orders of magnitude and can also probe PeV scale quarks. Searches for muon lepton flavor violation as well as precision measurements of $D^0$-$\bar D^0$ oscillations will also reach this interesting range. 

In Fig.~\ref{fig:Mh} we have assumed that the squark and slepton mass matrices are anarchic in flavor space. In contrast, the masses of the corresponding fermions are strongly hierarchical. In light of this we considered several possible models in which the fermion mass hierarchy is explained naturally and considered the impact on the sfermion flavor structure and the reach of low energy probes. Within models in which the fermion mass hierarchy is explained by flavor textures (e.g. extra dimensions, horizontal symmetries, etc) the structure of the sfermion mass matrix can fall into several categories, ranging from fully anarchical to hierarchical. In many of these cases we find that low energy probes remain sensitive to scalars near the PeV scale.

A distinct possibility is that the quark masses are generated radiatively.
Since split SUSY allows for large flavor violation it introduces a model building opportunity to dynamically explain the hierarchical structure of the SM quark and lepton masses  
through the hierarchy of superpartner loops~\cite{ArkaniHamed:1996zw,DiazCruz:2000mn,Ferrandis:2004ng,Crivellin:2011sj,Graham:2009gr}. In this way it is straightforward to account for the mass of the up quark even within a minimal model, as pointed out recently~\cite{ArkaniHamed:2012gw} (doing the same for the down quark and the electron requires additional model building and may require additional vector like fields). 
This attractive possibility interplays with the flavor and EDM bounds in interesting ways, because it implies a \emph{lower bound} on the amount of flavor violation and also has implications for possible alignment of CP phases. For this reason we pay special attention to the case of radiative mass generation, treating it separately when needed.

There is a rich taxonomy of mini-split SUSY scenarios. The higgsinos can be either at the TeV or the PeV scale. Also, $\tan\beta$ can be either large or small.  Sleptons can be somewhat lighter than squarks, but do not have to be so. The splitting between the  lightest SUSY particle (LSP) -- typically a wino -- and the gluino is about an order of magnitude,
but can be smaller if the higgsinos are heavy and/or there are extra vector-like fields at a PeV~\cite{ArkaniHamed:2012gw}. In our analysis we will not make any particular choice for these issues and will try to address the various cases when relevant.
 
The paper is structured as follows. In Section~\ref{sec:setup} we explain the setup and some of the assumptions we make in deriving our bounds. We also survey the current status of the various experimental bounds and their prospects in the near future. In Section~\ref{sec:EDMs} we consider limits from neutron and electron EDM searches. We analyze separately the large and small $\mu$  scenarios because different diagrams dominate in each of the two cases. 
In Section~\ref{sec:osc} we consider the limits from meson oscillations and in Section~\ref{sec:LFV} the  limits from lepton flavor violating processes, including $\mu\to e\gamma$, $\mu\to e$ conversion, and $\mu\to 3e$. Section~\ref{sec:fermion-masses} is devoted to models that explain the fermion mass hierarchy and the implications of our bounds for these frameworks. We consider two broad classes of ideas --- flavor textures and generation of fermion masses by loops of superpartners. In section~\ref{sec:conclusion} we conclude. Appendix \ref{appendix} is devoted to the large-log resummation, and Appendix \ref{app:loopfunctions} collects loop functions entering the $\mu\to e\gamma$ and $\mu\to e$ conversion predictions.

\section{The Setup and Main Highlights}\label{sec:setup}

We are interested in the supersymmetric spectra where the gauginos -- bino, wino and gluino -- are all at ${\mathcal O}({\rm TeV})$ scale, while sfermions -- squarks and sleptons -- are significantly heavier, with masses of ${\mathcal O}(10^2 {\rm ~TeV})- {\mathcal O}(10^3 {\rm ~TeV})$. Higgsinos could be as light as the gauginos or as heavy as the sfermions and we will consider these two cases separately when it makes a difference. For concreteness, we assume the MSSM field content. The mini-split SUSY spectrum means that it may be possible to observe gauginos at the LHC \cite{ArkaniHamed:2012gw}. However, the squarks and sleptons can only be probed through their virtual corrections to low energy processes. The sensitivity is due to the soft sfermion masses and trilinear couplings that act as new sources of flavor and CP violation.

Note that for PeV sfermions the left-right sfermion mixing is suppressed by ${\mathcal O}(m_f/m_{\tilde f})$ compared to the diagonal $m_{\tilde f}^2$, and can be neglected.
We do not make any assumptions about the flavor structure of the sfermion mass matrices, and thus parametrize the soft masses of squarks as 
\begin{equation}
 m_Q^2 = m_{\tilde q}^2 ( 1\!\!1 + \delta_q^L) ,~~ m_U^2 = m_{\tilde u}^2 ( 1\!\!1 + \delta_u^R) ,~~ m_D^2 = m_{\tilde d}^2 ( 1\!\!1 + \delta_d^R) ,
\end{equation}
and soft masses of sleptons
\begin{equation}
 m_L^2 = m_{\tilde \ell}^2 ( 1\!\!1 + \delta_\ell^L),~~ m_E^2 = m_{\tilde e}^2 ( 1\!\!1 + \delta_\ell^R) ,
\end{equation}
where $\delta_A$ are dimensionless matrices that encode the flavor breaking and mass splittings, and whose elements are all allowed to be ${\mathcal O}(1)$.  
We do not expect a strong mass hierarchy among the squark and slepton masses and set $m_{\tilde q}^2 = m_{\tilde u}^2 = m_{\tilde d}^2$ and $m_{\tilde \ell}^2 = m_{\tilde e}^2$, for simplicity. We work in the mass-insertion approximation where $(\delta_A)_{ij}$ are treated as perturbations and only the diagrams with the lowest numbers of $(\delta_A)_{ij}$ insertions are kept (here $A$ is a super-index and denotes both $L,R$ superscript and $q,u,d,l$ subscript dependence). In the numerical examples below, we use the super-CKM basis, where quarks are in the mass-basis and squark fields are rotated by the same unitary matrices that diagonalize the quarks. In the plots, we then always set the off-diagonal elements to $(\delta_A)_{ij}=0.3$. Incidentally, note that there is a relation between left-left down-squark and up-squark matrices,  $\delta_u^L = V \delta_d^L V^\dagger$, with $V$ the CKM matrix, so that $\delta_d^L = \delta_u^L ( 1 + {\mathcal O}(\lambda) )$ with $\lambda \simeq 0.23$ the sine of Cabibbo 
angle.

\begin{table}
\begin{center}
\begin{tabular}{cll}
\hline\hline
process & current exp.  & ~~future exp.~~  \\
\hline
$K^0$ mixing & $\epsilon_K = (2.228 \pm 0.011)\times 10^{-3}$~\cite{Beringer:1900zz}& --- \\
\multirow{2}{*}{$D^0$ mixing~} & ~~\multirow{2}{*}{$A_\Gamma = (-0.02 \pm 0.16)\% $~\cite{Amhis:2012bh}}~ & $\pm 0.007 \%$  LHCb~\cite{Bediaga:2012py} \\
 &  &  $\pm 0.06\%$  Belle II~\cite{Aushev:2010bq} \\
\multirow{2}{*}{$B_d$ mixing~} & \multirow{2}{*}{$\sin2\beta = 0.68 \pm 0.02$~\cite{Amhis:2012bh}} & $\pm 0.008 $  LHCb~\cite{Bediaga:2012py} \\
 &  &  $\pm 0.012 $  Belle II~\cite{Aushev:2010bq} \\
$B_s$ mixing & $\phi_s= 0.01 \pm 0.07$~\cite{Aaij:2013oba} & $\pm 0.008 $~LHCb~\cite{Bediaga:2012py} \\
\hline
$d_\text{Hg}$ & ~$< 3.1\times 10^{-29}~e{\rm cm}$~\cite{Griffith:2009zz} & $-$   \\
$d_\text{Ra}$ &  $-$   & ~$\lesssim 10^{-29}~e{\rm cm}$~\cite{Hewett:2012ns} \\
$d_n$ & $< 2.9\times 10^{-26} ~e{\rm cm}$~\cite{Baker:2006ts} & ~$\lesssim 10^{-28}~e{\rm cm}$~\cite{Hewett:2012ns} \\
$d_p$ & $-$ & ~$\lesssim 10^{-29} ~e{\rm cm}$~\cite{Hewett:2012ns} \\
$d_e$ & $< 1.05\times 10^{-27}~e{\rm cm}$~YbF~\cite{Hudson:2011zz,Kara:2012ay}~~ & ~$\lesssim 10^{-30} ~e{\rm cm}$~YbF, Fr~\cite{Hewett:2012ns} \\
& {$< 8.7 \times 10^{-29}~e{\rm cm}$~ ThO~\cite{Baron:2013eja}} ~~ & \\
\hline
$\mu \to e \gamma$ & $< 5.4 \times 10^{-13}$~MEG~\cite{Adam:2013mnn} & $\lesssim 6 \times 10^{-14}$~MEG upgrade~\cite{Baldini:2013ke} \\
$\mu \to 3 e$ & $ < 1.0 \times 10^{-12}$~SINDRUM I~\cite{Bellgardt:1987du} & $\lesssim 10^{-16}$~Mu3e~\cite{Blondel:2013ia} \\
~~$\mu \to e$ in Au~~ & $< 7.0 \times 10^{-13}$~SINDRUM II \cite{Bertl:2006up} &  $-$   \\
~~$\mu \to e$ in Al~~ &  $-$   & $\lesssim 6 \times 10^{-17}$~Mu2e~\cite{Abrams:2012er} \\
\hline\hline
\end{tabular}
\end{center}
\caption{\small Summary of current and selected future expected experimental limits on CP violation in meson mixing, EDMs and lepton flavor violating processes. 
}
\label{tab:experiments}
\end{table}

In the rest of the paper, we perform a detailed analysis of the impact that different low energy experimental results have on the allowed parameter space. The most important current and projected future limits on low energy probes that we use are collected in Tab.~\ref{tab:experiments}. For reader's convenience we also summarize the sensitivity of the considered low energy probes on the sfermion mass scale in Fig.~\ref{fig:Mh}.
Shown are the current and expected future constraints on the sfermion mass scale and $\tan\beta$ for the example of $|m_{\tilde W}|=|m_{\tilde B}|=$~3~TeV, $|m_{\tilde g}|=$~10~TeV and mass degenerate squarks,  sleptons and higgsinos, while taking the mass insertion parameters in the super-CKM basis to be $|(\delta_A)_{ij}| = 0.3$. All phases are assumed to be ${\mathcal O}(1)$ and chosen such that no large cancellations appear among the various SUSY contributions to the considered processes. In the case of meson mixing and EDMs, the constraints scale approximately as $\propto |(\delta_A)_{ij}|$, while in the case of the $\mu \to e$ transitions the scaling is approximately $\propto |(\delta_A)_{ij}|^{1/2}$. The dependence on the gaugino and the higgsino masses is discussed in detail below, in Secs.~\ref{sec:EDMs},~\ref{sec:osc}, and~\ref{sec:LFV}. The blue region preferred by a Higgs mass of $m_h = (125.5 \pm 1.0)$~GeV is obtained using 2-loop renormalization group running and 1-loop threshold corrections 
given in~\cite{Giudice:2011cg} and includes uncertainties from the top pole mass $m_t = 173.2 \pm 0.7$~GeV~\cite{Lancaster:2011wr} and the strong coupling constant $\alpha_s(m_Z) = 0.1184 \pm 0.0007$~\cite{Bethke:2009jm}. Note that for the higgs mass band we have taken the squarks to be universal. This assumption is obviously not satified in our framework. However, we have checked that the leading effects of flavor violation correspond to taking $m_{\tilde q}$  to be the weighted average of all squark mass eigenstates. Generically this will shift the blue band horizontally by an order one factor, which is a small effect on a log plot. 

Currently, CP violation in kaon mixing, given by $\epsilon_K$, is the only observable that is sensitive to squarks with PeV masses. Current constraints on CP violation in charm mixing and hadronic EDMs can probe squarks with 100 TeV masses.
Constraints on the slepton masses are still relatively modest in comparison, reaching tens of TeV.
In the future, EDMs of hadronic systems, such as the neutron and proton EDMs  are also projected to be able to probe CP violation induced by the PeV scale squarks. Projected improvements of the results on the electron EDM and $\mu \to e$ conversion in nuclei will be sensitive to sleptons at around 100 TeV and above. 

The phenomenology of PeV sfermions has recently been considered also in Refs. \cite{Moroi:2013vya,McKeen:2013dma,Moroi:2013sfa,Eliaz:2013aaa} (see also \cite{Sato:2013bta} for LHC implications and \cite{Dine:2013nga,Hisano:2013exa} for implications of proton decay bounds). Our work adds several new observables and aims for a comprehensive study. In particular, in the calculation of (C)EDMs we resum factors of $\log(|m_{\tilde g}|^2/m_{\tilde q}^2)$, as required because there is a large hierarchy between the gluino and squark mass scales.  The details of the resummation are given in Appendix~\ref{appendix}. We also consider simultaneously both one-loop and two loop contributions to EDMs. In our discussion we pay special attention to the scenario in which the up-quark mass is generated radiatively and the role that the neutron EDM plays in constraining this framework. Our work includes a study of meson mixing observables, including the promising prospects for $D-\bar D$ mixing.
We also consider the impact of low energy constraints on models of fermion masses.

\section{Electric Dipole Moments}\label{sec:EDMs}

Electric dipole moments (EDMs) of quarks and leptons, $d_{q,\ell}$, and chromo-electric dipole moments (CEDMs) of quarks, $\tilde d_q$, are described by dimension 5 operators in the effective Lagrangian,
\begin{equation} \label{eq:EDMs_Heff}
{\cal L}_{\rm eff} = -i \frac{d_\ell}{2} (\bar \ell \sigma^{\mu\nu} \gamma_5 \ell) F_{\mu\nu} -i \frac{d_q}{2}(\bar q \sigma^{\mu\nu} \gamma_5 q)F_{\mu\nu}  -i \frac{\tilde d_q}{2} g_s (\bar q  T^A\sigma^{\mu\nu} \gamma_5 q)G_{\mu\nu}^A ~.
\end{equation}
Note that we do not need to consider the CP violating dimension 6 operators -- the Weinberg 3 gluon operator~\cite{Weinberg:1989dx}, and CP violating 4 fermion operators~\cite{Barr:1991yx} -- since these are negligible in mini-split SUSY.

Currently, there are strong experimental bounds on the neutron ($d_n$), mercury ($d_\text{Hg}$), and electron ($d_e$) EDMs, 
as listed in Tab.~\ref{tab:experiments}.
All these bound are expected to improve significantly in the future~\cite{Hewett:2012ns,Kronfeld:2013uoa}. For instance by {two orders of magnitude for the electron and neutron EDM}, while experiments with radon and radium may reach sensitivities that correspond to an improvement of at least two orders of magnitude over the current mercury EDM bounds. All these improvements offer excellent opportunities to probe SUSY at the PeV scale.

The bounds on the neutron and mercury EDMs imply bounds on the quark EDMs and  CEDMs through the following relations~\cite{Pospelov:2005pr,Raidal:2008jk} 
\begin{eqnarray}
 d_{\rm Hg} &\simeq& 7\cdot10^{-3} e \left[ \tilde d_u(\hat\mu_h) - \tilde d_d(\hat\mu_h) \right] + 10^{-2} d_e, \label{dHg}\\
 d_n &\simeq& (1.4 \pm 0.6) \left[ d_d(\hat\mu_h) -0.25 d_u(\hat\mu_h) \right] + (1.1 \pm 0.5) e \left[ \tilde d_d(\hat\mu_h) + 0.5 \tilde d_u(\hat\mu_h) \right], \label{dn}
\end{eqnarray}
where the quark (C)EDMs are evaluated at the hadronic scale $\hat\mu_h \simeq 1$~GeV, and we quote directly the numerical values of the hadronic matrix elements. The uncertainties in the numerical coefficients of $d_\text{Hg}$ are relatively large, and can even be a factor of a few (see, e.g., the discussion in \cite{Engel:2013lsa}). Keeping this in mind, we show in the plots below the exclusions for the central values of hadronic matrix elements in Eqs. \eqref{dHg}, \eqref{dn}. 

\subsection{EDMs at One Loop -- Large \boldmath $\mu$} 

\begin{figure}[tb] 
\includegraphics[width=0.3\textwidth]{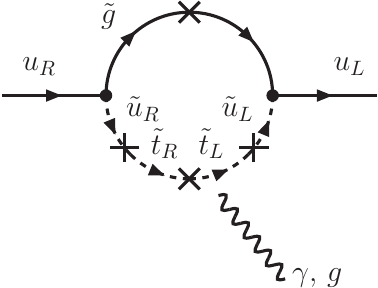}
\caption{\small An example of a flavor violating contribution to the up quark (C)EDM which dominates in mini-split SUSY for a $\mu$ term not much smaller than the squark masses.  The electron EDM is dominated by an analogous contribution with a gluino replaced by a bino. In the diagram, the photon (gluon) line can attach to any (color-)charged internal line.}
\label{fig:EDM_diagrams}
\end{figure}

When the $\mu$ parameter is large, $|\mu| \sim m_{\tilde q,\tilde \ell}$, the most important contributions to the (C)EDMs of light quarks and leptons come from the flavor violating diagrams, such as the one given in Fig.~\ref{fig:EDM_diagrams}~\cite{Hisano:2008hn,Altmannshofer:2009ne,Altmannshofer:2010ad,McKeen:2013dma}. In the flavor conserving case, the (C)EDMs of quarks and leptons are proportional to the masses of the corresponding quarks or leptons. With generic flavor mixing, however, the chirality flip can occur on the third generation sfermion line, enhancing the quark (C)EDMs by $m_{t}/m_{u}$ or $m_{b}/m_{d}$ and the electron EDM by $m_\tau/m_{e}$.\footnote{The anomalous magnetic moment of the electron, $\Delta a_e$, is similarly $m_\tau/m_e$ enhanced. However, for the current experimental bounds and the precision of the SM prediction~\cite{Giudice:2012ms}, $\Delta a_e$ typically gives much  weaker constraints than EDMs.} 
Assuming a common soft mass for the squarks $m_{\tilde q}$ and sleptons $m_{\tilde \ell}$ one arrives at
\begin{eqnarray} \label{eq:EDM_e}
\frac{d_e}{e} &=& \frac{\alpha_1}{4\pi} \frac{m_\tau}{m^2_{\tilde \ell} } \frac{|\mu m_{\tilde B}|}{m^2_{\tilde \ell}} t_\beta ~|\delta^L_{e\tau} \delta^R_{\tau e}|~ \frac{1}{2} \sin\phi_e ~, \\ \label{eq:EDM_u}
\left\{ \frac{d_u(m_{\tilde q})}{e} , \tilde d_u(m_{\tilde q}) \right\} &=& \frac{\alpha_s}{4\pi} \frac{m_t}{m_{\tilde q}^2} \frac{|\mu m_{\tilde g}|}{m_{\tilde q}^2} \frac{1}{t_\beta} ~|\delta^L_{ut} \delta^R_{tu}|~ \sin\phi_u ~\left\{ -\frac{8}{9} , - \frac{59}{6} - 3\log x \right\}~, \\ \label{eq:EDM_d}
\left\{ \frac{d_d(m_{\tilde q})}{e} , \tilde d_d(m_{\tilde q}) \right\} &=& \frac{\alpha_s}{4\pi} \frac{m_b}{m_{\tilde q}^2} \frac{|\mu m_{\tilde g}|}{m_{\tilde q}^2} t_\beta ~|\delta^L_{db} \delta^R_{bd}|~ \sin\phi_d ~\left\{ \frac{4}{9} , - \frac{59}{6} - 3\log x \right\}~,
\end{eqnarray}
where $x = |m_{\tilde g}|^2 / m_{\tilde q}^2 \ll 1$, while $\phi_e = \arg(\mu m_{\tilde B} \delta^L_{e\tau} \delta^R_{\tau e})$, $\phi_d = \arg(\mu m_{\tilde g} \delta^L_{db}\delta^R_{bd})$, and $\phi_u = \arg(\mu m_{\tilde g} \delta^L_{ut} \delta^R_{tu})$ are the rephasing invariant CP violating phases.
The results are of leading order in the mass insertion approximation and of leading order in $|m_{\tilde B}|^2 / m_{\tilde \ell}^2$ and $|m_{\tilde g}|^2 / m_{\tilde q}^2$ (expressions that hold also beyond the limit of large sfermion masses can be found in~\cite{Hisano:2008hn}).
Note that for the electron EDM, $d_e$, only the bino contribution is enhanced by $m_\tau/m_e$. The wino does not couple to the right-handed electron so that its contribution remains proportional to the electron mass, and is negligible. 

The quark CEDMs contain a logarithmically enhanced term, $\log(|m_{\tilde g}|^2/\tilde m^2)$, which is large because $|m_{\tilde g}| \ll m_{\tilde q}$. The large logarithm arises from a diagram in Fig. \ref{fig:EDM_diagrams} where the gluon attaches to the internal gluino line.
We resum the large log by performing a two step matching procedure: first integrating out squarks at the scale $\hat \mu=m_{\tilde q}$, calculating one-loop renormalization group (RG) running from $\hat \mu=m_{\tilde q}$ down to $\hat \mu=|m_{\tilde g}|$, and finally integrating out the gluino at the the scale $\hat \mu=|m_{\tilde g}|$, thus matching onto the usual (C)EDM effective Lagrangian \eqref{eq:EDMs_Heff}. 
The details of the resummation are relegated to Appendix~\ref{appendix}. 
The resummed versions of~(\ref{eq:EDM_u}) and~(\ref{eq:EDM_d}) for quark (C)EDMs, $d_{u,d}(|m_{\tilde g}|), \tilde d_{u,d}(|m_{\tilde g}|)$, are given in~(\ref{eq:CEDM_resummed}) and hold at the scale $\hat \mu=|m_{\tilde g}|$, where the gluino is integrated out. These are then evolved down to the hadronic scale $\hat\mu_h \simeq 1$~GeV using standard RG equations~\cite{Degrassi:2005zd}. Note that bino contributions to the quark CEDMs are suppressed by the small $U(1)$ gauge coupling and also do not contain the above log enhancement, and are therefore safely neglected.

\begin{figure}[tb] 
\centering
\includegraphics[width=0.46\textwidth]{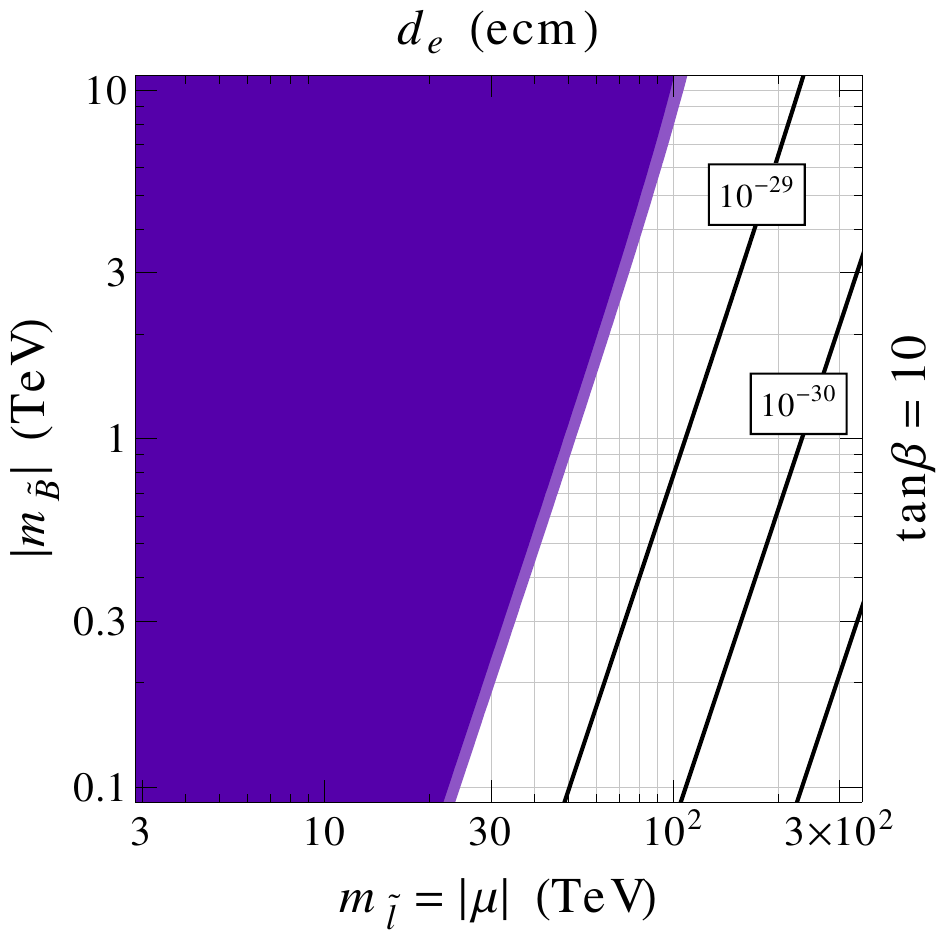}~~~~~~~~\includegraphics[width=0.46\textwidth]{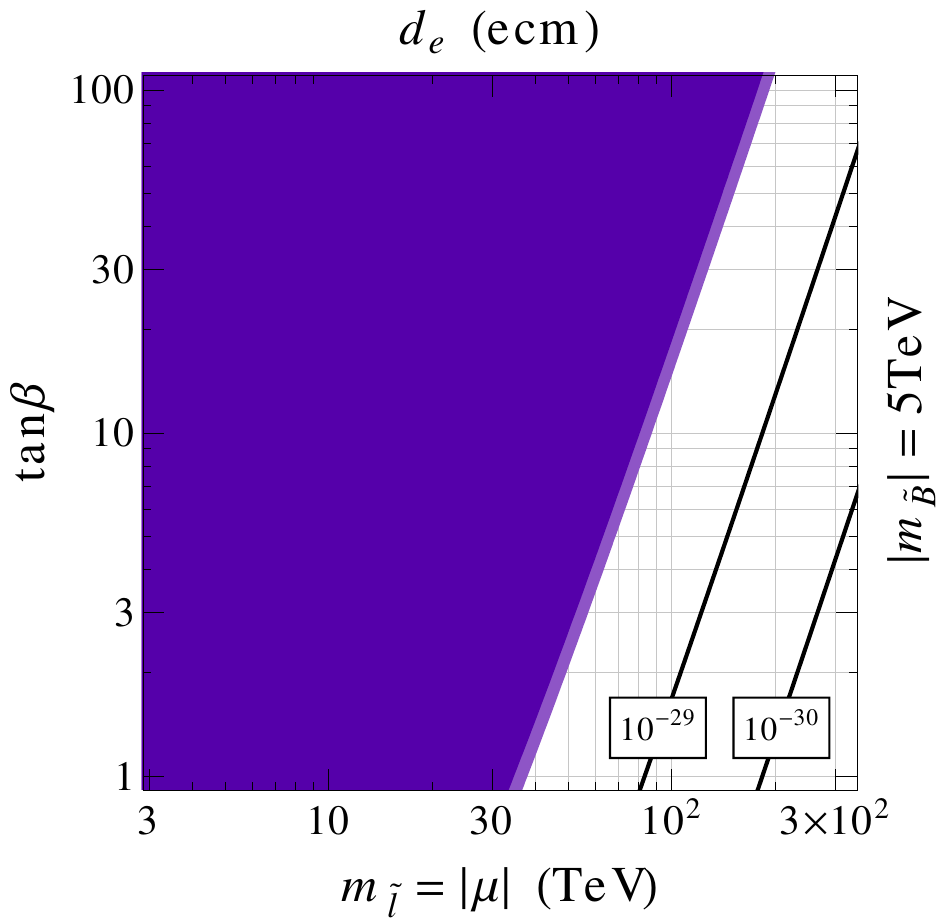}
\caption{\small
The  dark (light) shaded regions show 95\% (90\%) C.L. exclusions from the current electron EDM experiments. In the left and right plot, $\tan\beta=10$ and $|m_{\tilde B}| =$~5~TeV are held fixed, respectively. The relevant mass insertions are set to $(\delta_A)_{ij}=0.3$, and the CP violating phase to $\sin\phi_e=1$.
The solid lines show the predicted electron EDM values.
}
\label{fig:EDMs_1}
\end{figure}

\begin{figure}[tb] 
\centering
\includegraphics[width=0.46\textwidth]{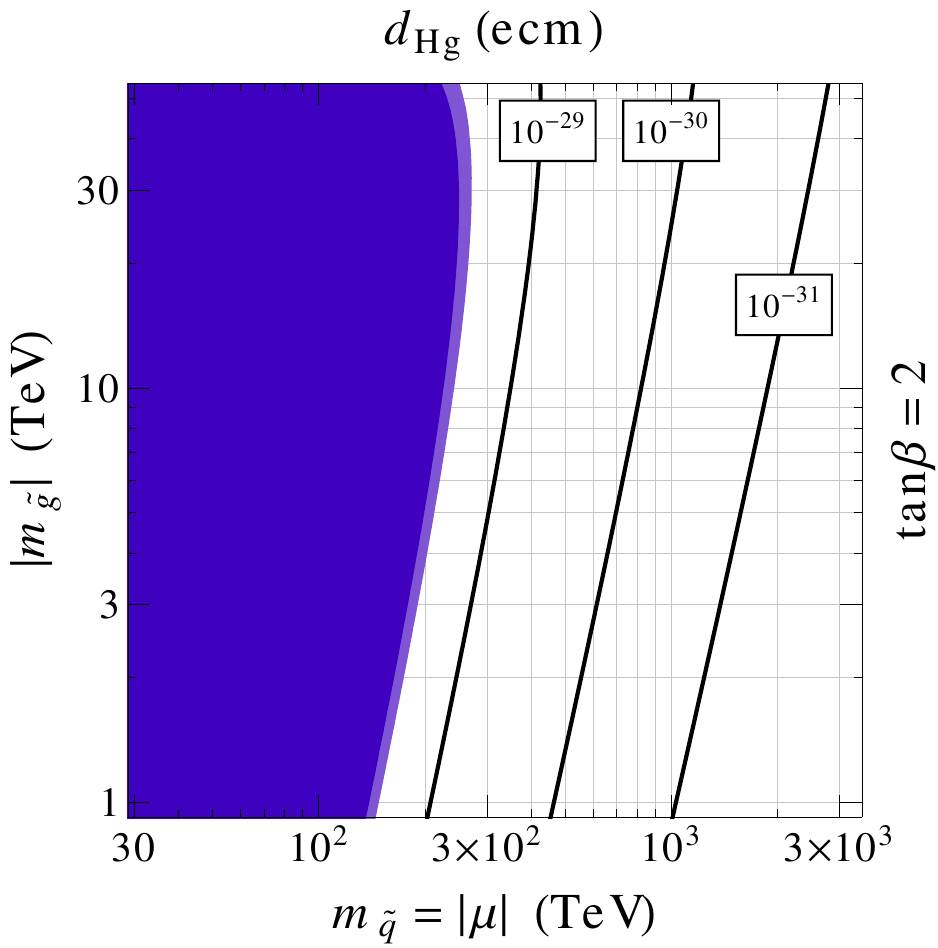}~~~~~~~~\includegraphics[width=0.46\textwidth]{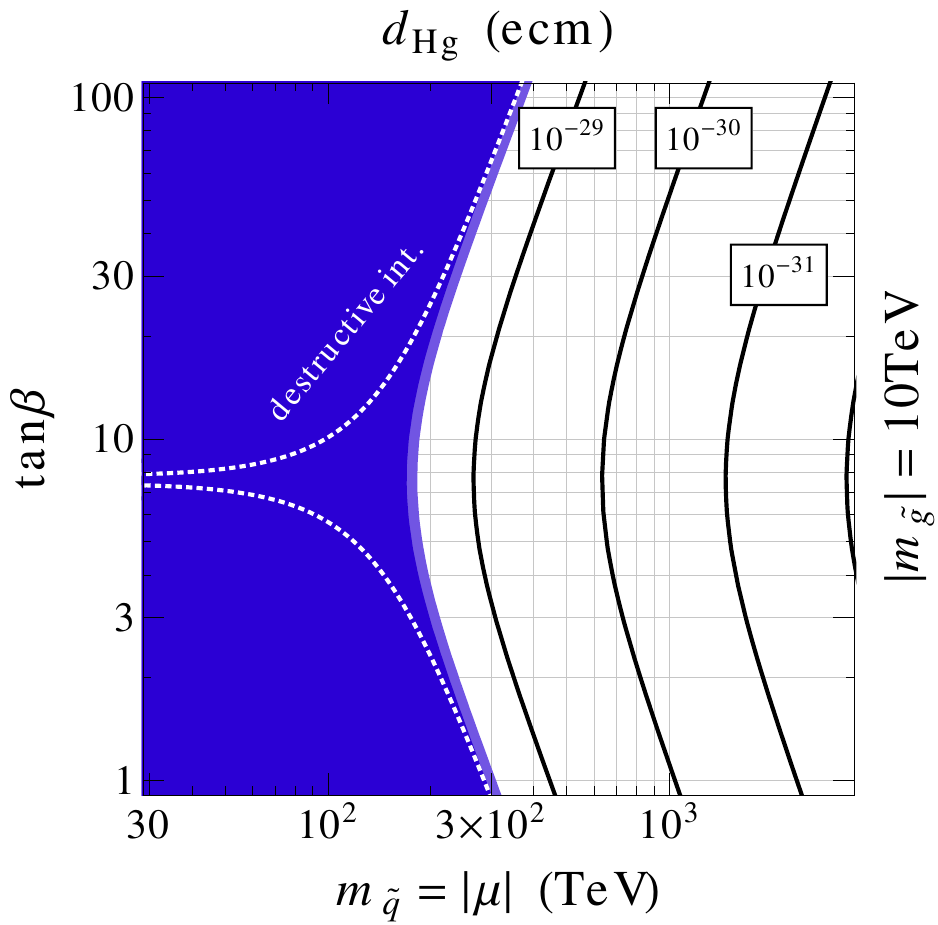}\\
\includegraphics[width=0.46\textwidth]{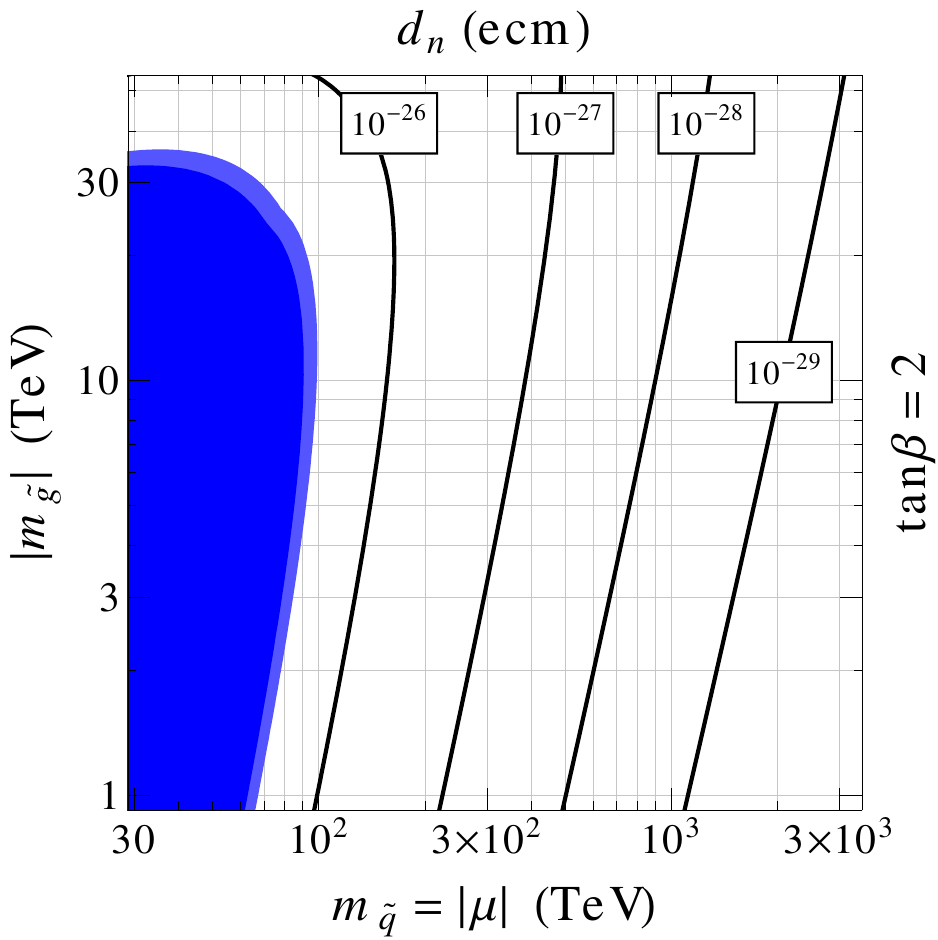}~~~~~~~~\includegraphics[width=0.46\textwidth]{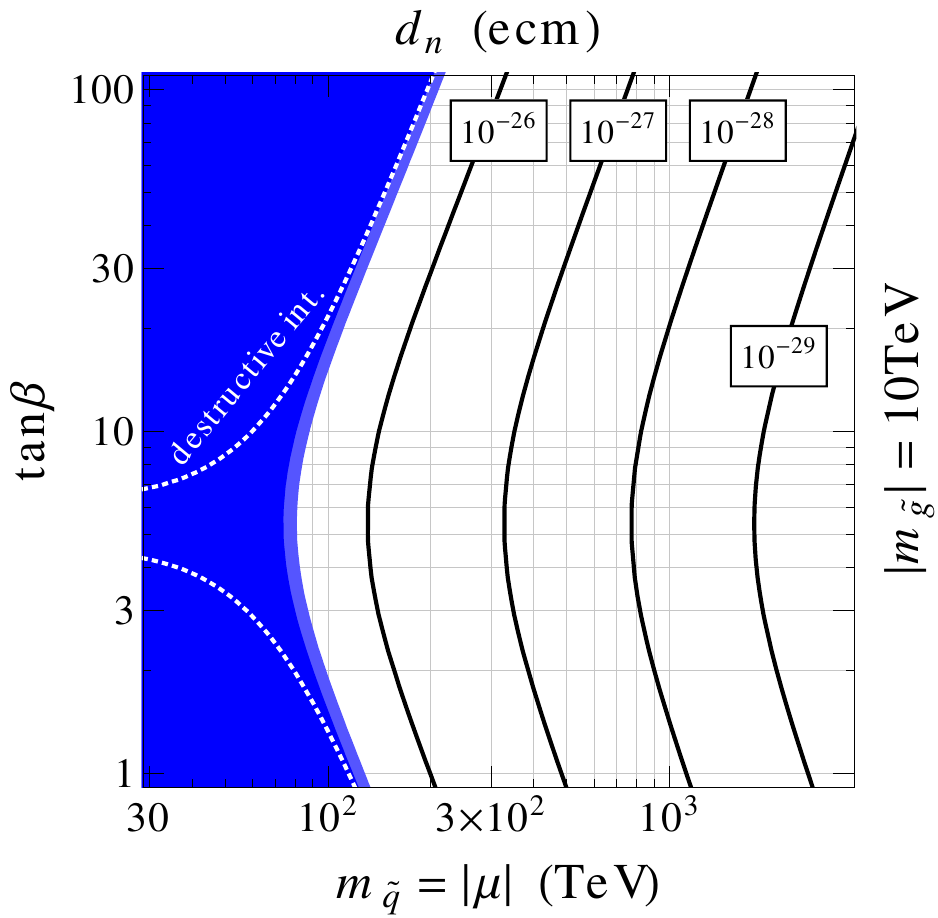}
\caption{\small
The dark (light) shaded regions show 95\% (90\%) C.L. exclusions from current neutron and mercury EDM experiments assuming constructive interference of up and down quark contributions (dotted white lines denote 95\% C.L. limits for destructive interference). In the left and right columns, $\tan\beta=2$ and $|m_{\tilde g}|=$~10~TeV are held fixed for mercury and neutron EDM, from top to bottom. The relevant mass insertions are set to $(\delta_A)_{ij}=0.3$, and the CP violating phases to $\sin\phi_q=1$.
The solid lines show the predicted EDM values.
}
\label{fig:EDMs_2}
\end{figure}

We next turn to the numerical evaluation of the above expressions and the impact current and future limits on EDMs have on 
the mini-split SUSY parameter space. We are first interested in the case where the $\mu$ parameter is of order the sfermion masses, which means that the one loop flavor violation enhanced contributions \eqref{eq:EDM_e}, \eqref{eq:EDM_u}, \eqref{eq:EDM_d} dominate. Up to resummation effects, the bounds are stronger, if the gluino is heavier (or bino for electron EDM). The bounds are also linear in the $\mu$ parameter. The constraints, on the other hand, become irrelevant, if either the flavor violating mass insertions $(\delta_A)_{ij}$ or the CP violating phases $\phi_A$ are small. For the numerical examples in Figs.~\ref{fig:EDMs_1} and~\ref{fig:EDMs_2}, we keep the $\mu$ parameter equal to the squark mass, $|\mu|=m_{\tilde q}$, for mercury and neutron EDMs (or slepton mass $|\mu|=m_{\tilde \ell}$ for the electron EDM). All relevant mass insertion parameters are taken to be $|\delta_{ij}^L| = |\delta_{ij}^R| = 0.3$ and CP violating phases of $\sin\phi_i = 1$ are assumed in all cases. The dark (light) shaded 
regions are excluded at the 95\% (90\%) C.L. by current measurements, assuming constructive interference between up and down quark contributions to the neutron and mercury EDM. The dotted white lines show the 95\% C.L. limits in the case of destructive interference. The left columns of plots in Figs.~\ref{fig:EDMs_1} and~\ref{fig:EDMs_2} illustrate the increased relevance of EDM bounds for larger gaugino masses. The right columns on the other hand illustrate the $\tan\beta$ $(\tan\beta^{-1})$ behavior of $\tilde d_d$ $(\tilde d_u)$, which dominate in the bounds from mercury and neutron EDM for large (small) $\tan\beta$. The dependence of the electron EDM on $\tan\beta$, on the other hand, is always linear.  Note also, that the exclusions become more stringent with growing $m_{\tilde g}$, as long as $m_{\tilde g}\ll m_{\tilde q}$. This increase saturates for $m_{\tilde g}\sim m_{\tilde q}$, as illustrated in bottom left panel of Fig. \ref{fig:EDMs_2} (incidentally in this region also the expanded expressions \eqref{eq:EDM_u}, \eqref{eq:EDM_d} are no longer valid).

From Fig. \ref{fig:EDMs_1} we also see that the current limit on the electron EDM probes slepton masses of {${\mathcal O}$(30 TeV) (small $\tan\beta\simeq1$) to ${\mathcal O}$(150 TeV) (large $\tan\beta\simeq30$)}. Future sensitivities for the electron EDM at the level of $10^{-30} \,e\,$cm will allow to probe slepton masses of several 100's of TeV (small $\tan\beta$) and even beyond 1000 TeV (large $\tan\beta$).
Current limits on the neutron and mercury EDM can already test squark masses above 100~TeV both for small and large $\tan\beta$.
Improving sensitivities by 2 orders of magnitude will allow to probe squarks at the PeV scale and above.

\subsection{EDMs at Two Loops - Small \boldmath $\mu$}\label{sec:EDM2loop}

\begin{figure}[tb] 
 \includegraphics[width=0.33\textwidth]{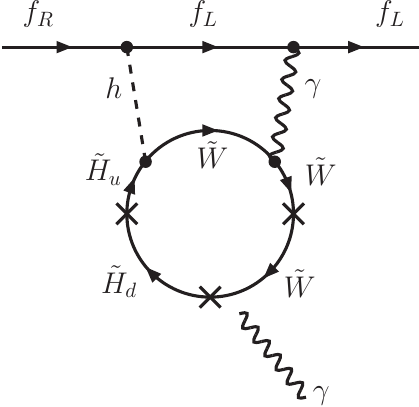}
\caption{\small Example of a 2-loop Barr-Zee contribution to fermion EDMs which dominates in mini-split SUSY for small values of the $\mu$ term.}
\label{fig:EDM_diagram-BarrZee}
\end{figure}

Next, let us consider the case of mini-split SUSY where the $\mu$ parameter is small, $|\mu| \ll m_{\tilde q, \tilde \ell}$. In this case, 2-loop Barr-Zee type diagrams containing light charginos can give the dominant contributions to the EDMs~\cite{Giudice:2005rz}. An example diagram is shown in Fig. \ref{fig:EDM_diagram-BarrZee}. 
These 2-loop contributions,  can be important only if the $\mu$ term is small, of the order of the wino mass, such that the charginos have a non-negligible higgsino component. For higher $\mu$ values the chargino-higgsino mixing angle scales as $\sim m_W/\mu$, so that the Barr-Zee type 2-loop contributions quickly decouple as $\mu^{-1}$. This has to be contrasted with the 1-loop flavor violation enhanced contributions to EDMs discussed above, that grow linearly with the $\mu$ parameter. Combining both types of contributions, EDMs can probe complementary regions of parameter space. 

\begin{figure}[tb] \centering
\includegraphics[width=0.96\textwidth]{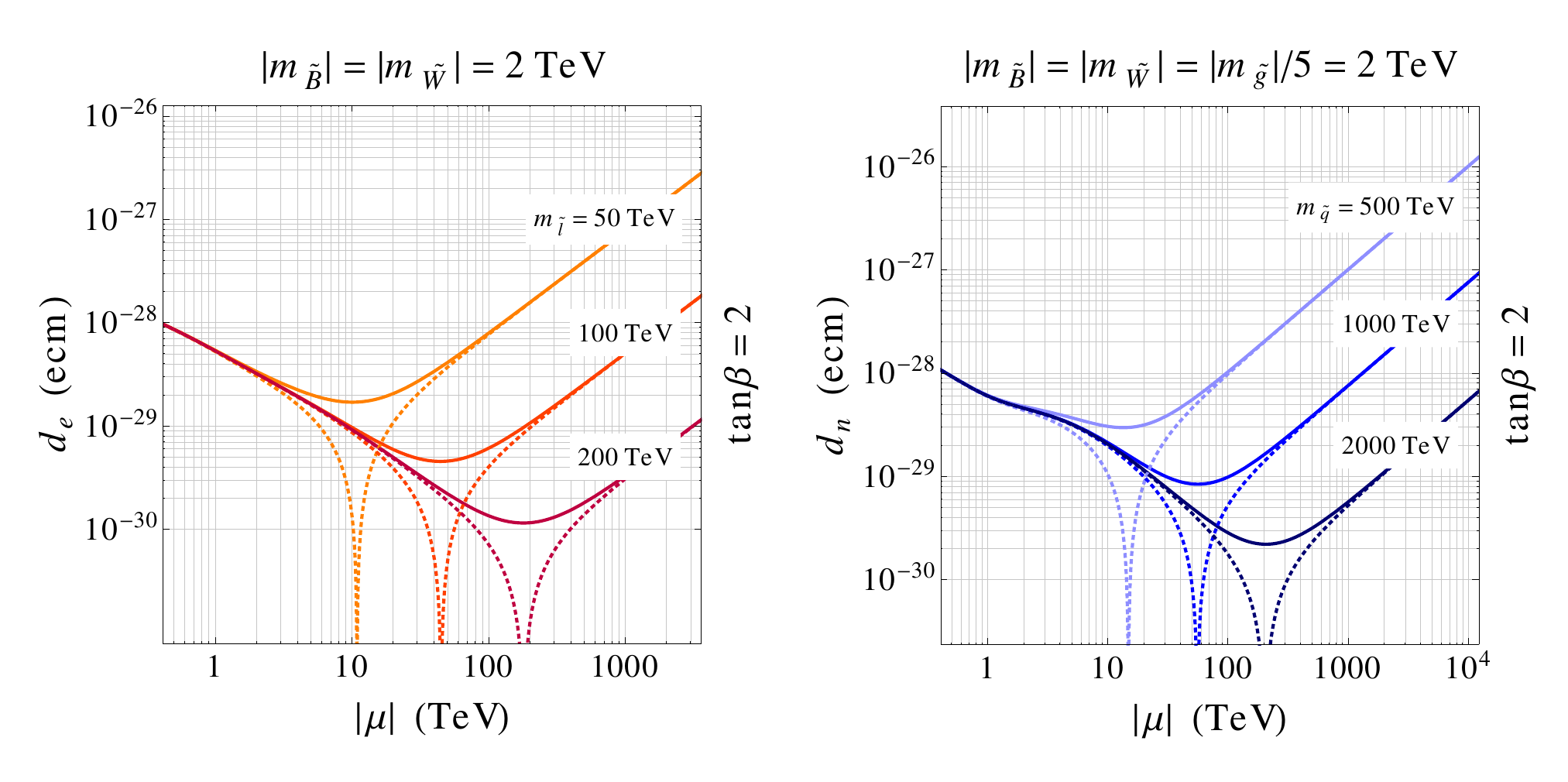}
\caption{\small
The electron EDM (left plot) and neutron EDM (right plot) as function of the higgsino mass $|\mu|$ for various values of the slepton masses $m_{\tilde \ell}$ and squark masses $m_{\tilde q}$, as indicated. We set $\tan\beta = 2$ and fix gaugino masses to $|m_{\tilde B}| = |m_{\tilde W}| = 2$~TeV and $|m_{\tilde g}| = 10$~TeV. All mass insertion parameters are taken to be $|\delta_{ij}^L| = |\delta_{ij}^R| = 0.3$ and CP violating phases of $\sin\phi_i = 1$ are assumed. The solid (dashed) lines correspond to constructive (destructive) interference between 1-loop and 2-loop contributions to the EDMs.}
\label{fig:EDMs_mu}
\end{figure}

This is illustrated in Fig.~\ref{fig:EDMs_mu}, which shows the electron EDM (left plot) and the neutron EDM predictions (right plot)  as a function of the $\mu$ parameter, for several values of the slepton masses $m_{\tilde \ell}$ and squark masses $m_{\tilde q}$ and choosing a small $\tan\beta = 2$, as indicated. 
In both plots, the gaugino masses are fixed to exemplary values $|m_{\tilde B}| = |m_{\tilde W}| = 2$~TeV, $|m_{\tilde g}| = 10$~TeV, 
mass insertion parameters are taken to be $|\delta_{ij}^L| = |\delta_{ij}^R| = 0.3$ and CP violating phases of $\sin\phi_i = 1$ are assumed. 
The solid (dashed) lines assume constructive (destructive) interference between the 1-loop and 2-loop contributions to the EDMs. For $|\mu|\sim {\mathcal O}(1{\rm TeV})$ the predicted EDMs are within reach of future sensitivities, and are independent of sfermion masses, because the 2-loop contribution dominates. 
The EDM predictions do depend on the sfermion masses for large values of $|\mu|$. For $|\mu| = {\mathcal O}(10^3 \text{~TeV})$ and the chosen small $\tan\beta = 2$, slepton masses of $m_{\tilde \ell} = {\mathcal O}(100 \text{~TeV})$ and squark masses of  $m_{\tilde q} = {\mathcal O}(10^3 \text{~TeV})$ are within reach of future EDM experiments. We have verified that the addition of A-terms with a size expected from anomaly mediation (loop suppressed) does not change Fig.~\ref{fig:EDMs_mu} significantly.

\section{Meson Oscillations} \label{sec:osc}

Meson oscillations, especially kaon mixing, are known to be highly sensitive probes of new sources of quark flavor violation. 
New physics with generic flavor structure that contributes to kaon mixing at tree level is constrained up to scales of $\sim 10^5$ TeV~\cite{Isidori:2010kg}. In our setup, contributions to meson oscillations arise only at the loop level. Nonetheless, very high SUSY scales of $\sim 10^3$ TeV can be probed.

Short distance contributions to meson oscillations are described by an effective Hamiltonian
\begin{equation} \label{eq:mixingq_Heff}
\mathcal{H}_{\rm eff} = \sum_{i=1}^5 C_i Q_i + \sum_{i=1}^3 \tilde C_i \tilde Q_i ~+{\rm h.c.}~,
\end{equation}
where the dimension 6 operators most relevant for the case of mini-split SUSY are
\begin{eqnarray}
Q_1 = (\bar u^\alpha \gamma_\mu P_L c^\alpha)(\bar u^\beta \gamma^\mu P_L c^\beta),~&&~~ \tilde Q_1 = (\bar u^\alpha \gamma_\mu P_R c^\alpha)(\bar u^\beta \gamma^\mu P_R c^\beta), \nonumber \\
Q_4 = (\bar u^\alpha P_L c^\alpha)(\bar u^\beta P_R c^\beta),~&&~~ Q_5 = (\bar u^\alpha P_L c^\beta)(\bar u^\beta P_R c^\alpha),
\end{eqnarray}
with $P_{R,L}=\frac{1}{2}(1\pm\gamma_5)$ and $\alpha,\beta$ are color indices, and we chose the quark flavors relevant
for $D^0 - \bar D^0$ mixing. The corresponding operators for kaon, $B_d$, and $B_s$ mixing are obtained by obvious replacements of the quark fields.

The dominant SUSY contributions to the Wilson coefficients $C_i$ come from box diagrams with gluino--squark loops. At the scale $\hat\mu = m_{\tilde q}$ where squarks are integrated out, we have to leading order in $|m_{\tilde g}|^2 / m_{\tilde q}^2$ and at leading order in the mass insertions\footnote{Mass insertion approximation expressions valid for $|m_{\tilde g}| \sim m_{\tilde q}$ can be found, e.g., in~\cite{Gabbiani:1996hi,Altmannshofer:2009ne}. The full set of MSSM 1-loop contributions in the mass eigenstate basis are collected in~\cite{Altmannshofer:2007cs}.}
\begin{eqnarray}
 C_1(m_{\tilde q}) = -\frac{\alpha_s^2}{m_{\tilde q}^2} (\delta^L_{cu})^2 \frac{11}{108},~&&~~ \tilde C_1(m_{\tilde q}) = -\frac{\alpha_s^2}{m_{\tilde q}^2} (\delta^R_{cu})^2 \frac{11}{108}, \\
 C_4(m_{\tilde q}) = \frac{\alpha_s^2}{m_{\tilde q}^2} (\delta^R_{cu} \delta^L_{cu}) \frac{1}{9},~&&~~ C_5(m_{\tilde q}) = - \frac{\alpha_s^2}{m_{\tilde q}^2}(\delta^L_{cu} \delta^R_{cu}) \frac{5}{27}.
\end{eqnarray}
Note that in the considered limit of $|m_{\tilde g}|^2 \ll m_{\tilde q}^2$ the Wilson coefficients are to an excellent approximation independent of the gluino mass and only depend on the mass scale of the squarks.
Using renormalization group equations, the Wilson coefficients can be evolved down to hadronic scales, where lattice evaluations of the matrix elements of the operators in~(\ref{eq:mixingq_Heff}) are given~\cite{Bertone:2012cu,Lubicz:2008am,Bouchard:2011xj,Carrasco:2013zta}.
The presence of a dynamical gluino below $m_{\tilde q}$ does not change the anomalous dimensions of the operators at leading order (LO)~\cite{Ciuchini:1997bw,Contino:1998nw,Buras:2000if,Kersten:2012ed}. Therefore, the only effect of the gluino on the running of the Wilson coefficients at LO comes from the slightly modified running of $\alpha_s$ that is given in the Appendix in~(\ref{eq:beta_s}).

We derive bounds on the squark scale by combining available experimental information on CP conserving and CP violating observables in meson mixing. In particular, for kaon mixing we consider the mass difference $\Delta M_K$~\cite{Beringer:1900zz} and the CP violating observable $\epsilon_K$~\cite{Beringer:1900zz}. 
In the case of $D^0$ mixing we use the combined experimental information on the normalized mass and width differences ($x$ and $y$) as well as the CP violating parameters $|q/p|$ and $\phi$. We assume the absence of direct CP violation in charm decays and take into account the full error correlation matrix provided by~\cite{Amhis:2012bh}. In the case of $B_d$ and $B_s$ mixing we consider the mass differences $\Delta M_d$~\cite{Amhis:2012bh} and $\Delta M_s$~\cite{Abulencia:2006mq,Aaij:2013mpa} as well as the information on the $B_d$ and $B_s$ mixing phases extracted from $B_d \to J/\psi K^0$~\cite{Amhis:2012bh} as well as $B_s \to J/\psi K^+K^-$ and $B_s \to J/\psi \pi^+\pi^-$~\cite{Aaij:2013oba}. 
For the SM predictions, we use CKM inputs from~\cite{Charles:2004jd}.

\begin{figure}[tb] \centering
\includegraphics[width=0.46\textwidth]{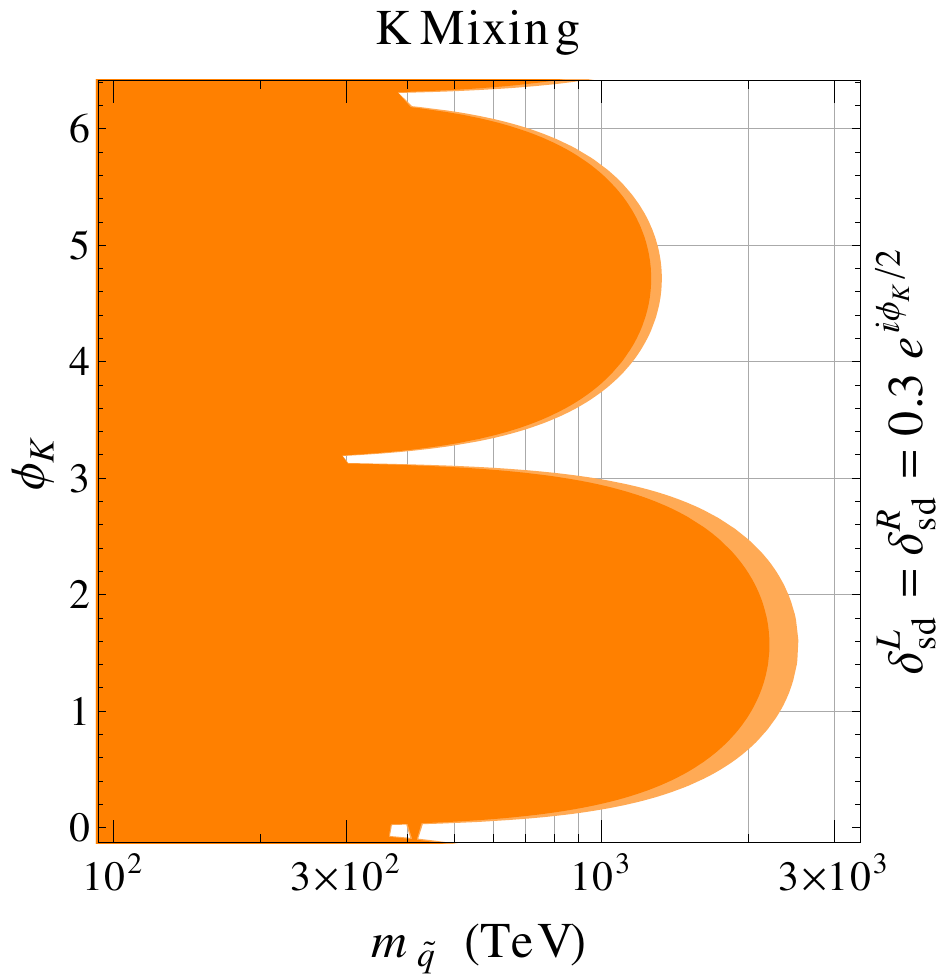} ~~~~~~~~~ \includegraphics[width=0.46\textwidth]{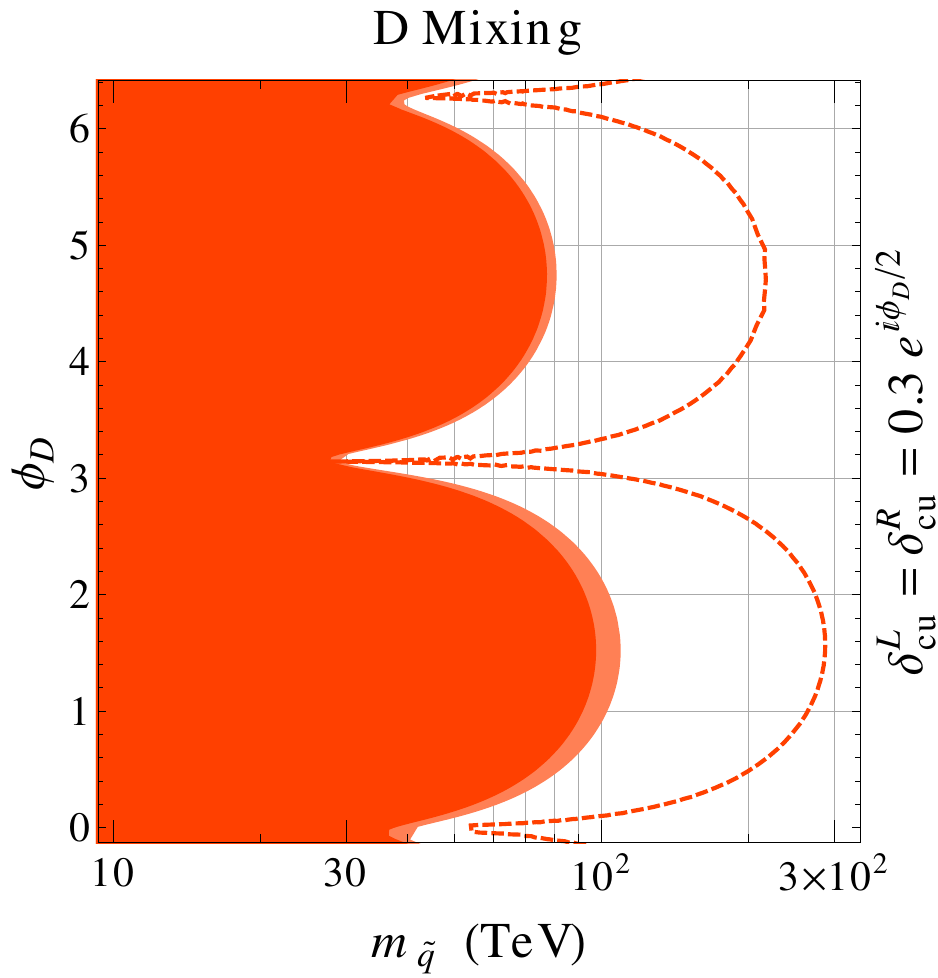} \\[10pt]
\includegraphics[width=0.46\textwidth]{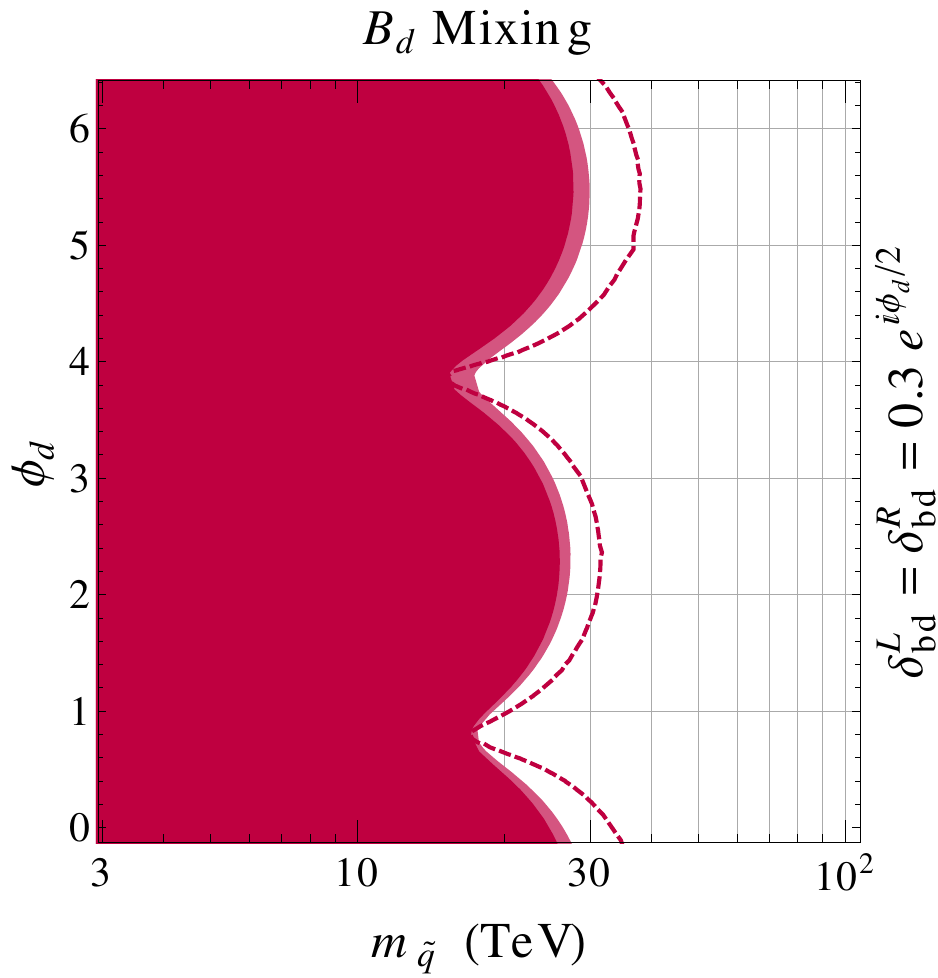} ~~~~~~~~~ \includegraphics[width=0.46\textwidth]{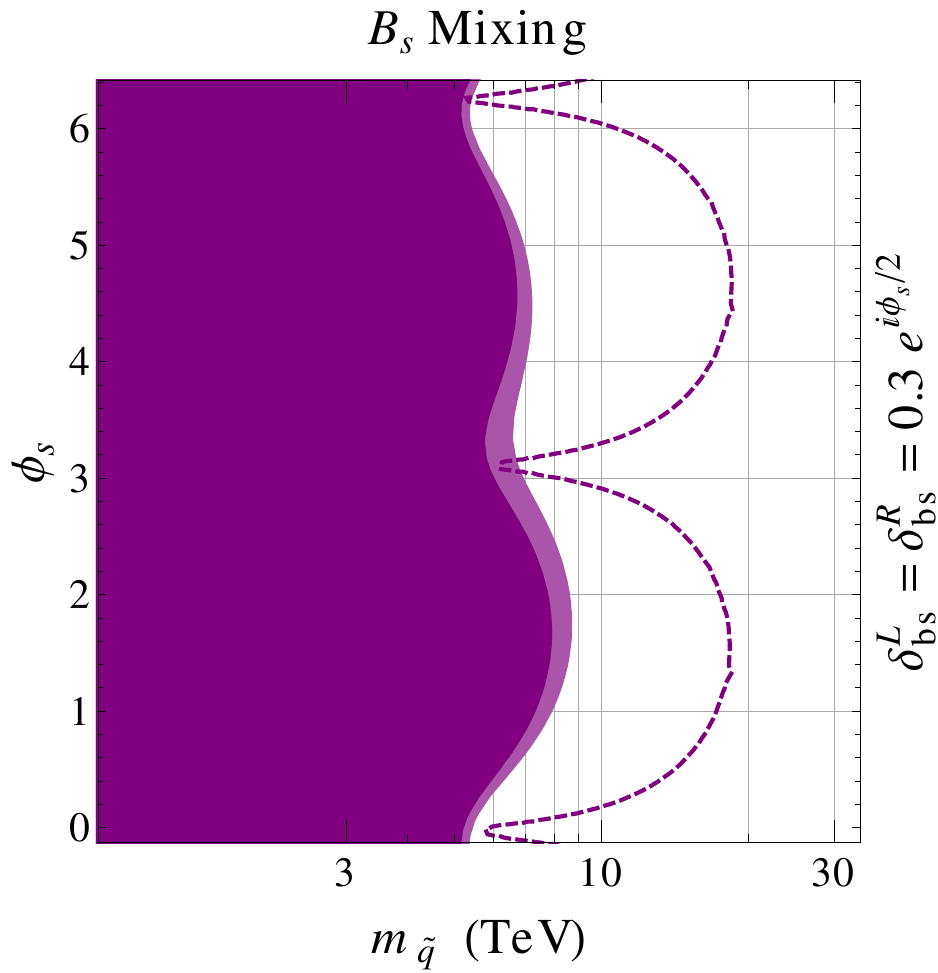}
\caption{\small
Squark masses $m_{\tilde q}$ probed by meson oscillations as a function of the phase of the NP contribution $\phi_i$. The gluino mass is fixed to $|m_{\tilde g}| = 3$~TeV. The dark (light) shaded regions are excluded at 95\% (90\%) C.L.. The dashed lines show the expected 95\% C.L. constraints with future experimental improvements on CP violation in meson mixing (factor $\sim 10$ in $D^0$ mixing, factor $\sim 2$ in $B_d$ mixing, and factor $\sim 10$ in $B_s$ mixing).
}
\label{fig:mixing}
\end{figure}

The resulting bounds on the squark mass scale are shown in Fig.~\ref{fig:mixing} as function of the phase of the NP contribution to the mixing amplitudes $\phi_i=\arg(M_{12,i}^\text{NP})$, $i = K, D, B_d, B_s$. The dark (light) shaded regions are excluded at 95\% (90\%) C.L.. In the plots we fix the mass insertions to moderate values of $\delta_{ij}^L = \delta_{ij}^R = 0.3 \times e^{i \phi_i/2}$. The gluino mass is fixed to $|m_{\tilde g}| = 3$~TeV. As already discussed above, the results are to a very good approximation independent of the gluino mass as long as $|m_{\tilde g}| \ll m_{\tilde q}$. 
In regions of parameter space that are probed by the current and future data, the NP effects in meson oscillations are dominated by contributions to the Wilson coefficients $C_4$ and $C_5$. At leading order in the mass insertions, these coefficients depend on the same combinations $(\delta_{ij}^L \delta_{ij}^R)$ and therefore the derived bounds are to a very good approximation independent of the relative phase between the $(\delta_{ij}^L)$ and $(\delta_{ij}^R)$.
The derived bounds do depend strongly on the overall phase of $(\delta_{ij}^L \delta_{ij}^R)$, especially in the case of kaon and $D^0$ mixing, where 
the constraints on CP violating NP contributions are considerably stronger than constraints on  the CP conserving ones.

Squark masses above 1000~TeV can be probed with kaon mixing, if the corresponding NP phase is ${\mathcal O}(1)$. The NP reach of kaon mixing is limited by the uncertainty on the SM predictions that are not expected to improve significantly in the foreseeable future. The current constraints from $D^0$ mixing reach up to $m_{\tilde q} \sim 100$~TeV as long as the NP phase is not accidentally suppressed. Future experimental bounds on CP violation in $D^0$ mixing at LHCb~\cite{Bediaga:2012py} and Belle~II~\cite{Aushev:2010bq} are expected to improve by at least one order of magnitude and can potentially probe scales of $\sim 300$~TeV as indicated by the dashed line in the upper right plot.
The scales that are currently probed by $B_d$ and $B_s$ mixing are much lower. Also with the expected improved measurements of the mixing phases at LHCb~\cite{Bediaga:2012py} and Belle~II~\cite{Aushev:2010bq}, only squark masses of $20 - 40$ TeV can be reached.

\section{Lepton Flavor Violation} \label{sec:LFV}

Another important set of constraints that is very sensitive to new particles and interactions beyond the SM are the 
charged lepton flavor violating (LFV) processes. We focus on $\mu\to e$ transitions which give by far the strongest constraint in the case of generic lepton flavor violation.
Current bounds on the $\mu \to e \gamma$ decay, the $\mu \to 3 e$ decay, and $\mu \to e$ conversion in nuclei probe NP up to masses of 1000 TeV, if NP is contributing 
at tree-level and has generic flavor violation~\cite{deGouvea:2013zba,Bernstein:2013hba}. Future sensitivities, with especially large improvement expected in $\mu \to e$ conversion sensitivity, may extend the reach above 10,000 TeV~\cite{Hewett:2012ns,deGouvea:2013zba,Kronfeld:2013uoa,Bernstein:2013hba}. In SUSY frameworks the above LFV processes are loop induced, so that slepton masses up to the PeV scale may be probed in the future. We focus on the most important contributions in mini-split SUSY, while the complete set of 1-loop contributions in the mass eigenstate basis for the LFV processes discussed in this section can be found in~\cite{Hisano:1995cp,Arganda:2005ji}. We will find that in split supersymmetry the dipole operator tends to dominate the rate for the two decay modes, but $\mu \to e$ conversion can be dominated by photon and $Z$ penguins. This should be contrasted with TeV SUSY, where the dipole tends to dominate for all three processes.

\subsection{The \texorpdfstring{\boldmath $\mu \to e \gamma$}{mu --> e gamma} Decay} \label{sec:muegamma}

The current bound on the branching ratio of the radiative $\mu \to e\gamma$ decay from the MEG experiment is~\cite{Adam:2013mnn}
\begin{equation} \label{eq:muegamma_exp}
 {\rm BR}(\mu \to e \gamma) < 5.7 \times 10^{-13} ~~~@~90\%~ \textnormal{C.L.}~.
\end{equation}
A MEG upgrade can improve this limit by another order of magnitude down to BR$(\mu \to e\gamma) \lesssim 6 \times 10^{-14} $~\cite{Baldini:2013ke}.

The $\mu \to e \gamma$ branching ratio can be written as
\begin{equation}
  {\rm BR}(\mu \to e \gamma) \simeq \frac{{\rm BR}(\mu \to e \gamma)}{{\rm BR}(\mu \to e \nu\bar\nu)} = \frac{48 \pi^3 \alpha_{\rm em}}{G_F^2} \Big( |A^L_{\mu e}|^2 + |A^R_{\mu e}|^2\Big) ~,
\end{equation}
where the amplitudes $A^i_{\mu e}$ are the coefficients of higher dimensional operators in an effective theory description of the decay
\begin{equation} \label{eq:Heff_muegamma}
 \mathcal{H_{\rm eff}} = e \frac{m_\mu}{2} \Big( A^L_{\mu e}~ \bar e \sigma^{\mu\nu} P_L \mu + A^R_{\mu e}~ \bar e \sigma^{\mu\nu} P_R \mu \Big)F_{\mu\nu} ~.
\end{equation}

\begin{figure}[tb] \centering
\includegraphics[width=0.29\textwidth]{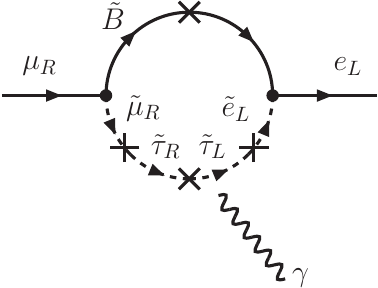} ~~~~~~~ \includegraphics[width=0.29\textwidth]{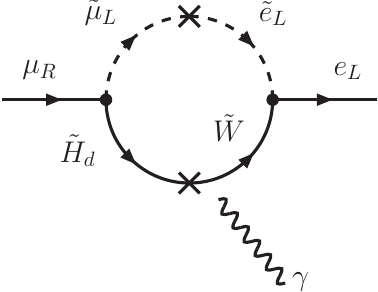} ~~~~~~~ \includegraphics[width=0.29\textwidth]{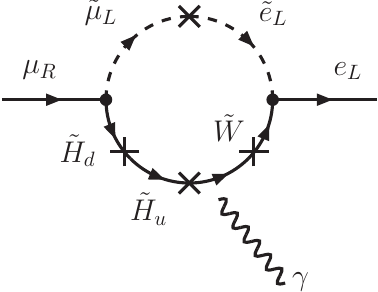}
\caption{\small Example contributions to the $\mu \to e \gamma$ amplitude. Left: flavor enhanced bino contribution. Middle and right: wino-higgsino contributions.}
\label{fig:muegamma}
\end{figure}

In mini-split SUSY, the most important contributions to the $\mu \to e \gamma$ amplitude arise from bino and wino loops~\cite{Hisano:1995cp,Arganda:2005ji,Paradisi:2005fk}. Higgs mediated contributions to $\mu \to e \gamma$ can be very important in TeV scale SUSY with large $\tan\beta$~\cite{Paradisi:2006jp}, but are negligible in mini-split SUSY. The dominant bino contribution arises at second order in mass insertions, ${\mathcal O}(\delta^R \delta^L)$, and involves mixing into the third generation which leads to an enhancement factor of $m_\tau / m_\mu$ over the contributions linear in the mass insertions. The relevant Feynman diagram is shown in Fig.~\ref{fig:muegamma} (left-most diagram) and gives
\begin{eqnarray} \label{eq:AL}
A_{\mu e}^{L,\tilde B} &=& \frac{\alpha_1}{4\pi} \left(\frac{m_\tau}{m_\mu}\right) \frac{\mu m_{\tilde B}}{m_{\tilde \ell}^4} ~t_\beta~ \frac{1}{2} (\delta^R_{\mu\tau} \delta^L_{\tau e}) ~, \\ \label{eq:AR}
A_{\mu e}^{R,\tilde B} &=& \frac{\alpha_1}{4\pi} \left(\frac{m_\tau}{m_\mu}\right) \frac{\mu m_{\tilde B}}{m_{\tilde \ell}^4} ~t_\beta~ \frac{1}{2} (\delta^L_{\mu\tau} \delta^R_{\tau e}) ~.
\end{eqnarray}
The expressions hold in the limit $|m_{\tilde B}| \ll m_{\tilde \ell}$. The bino contributions \eqref{eq:AL}, \eqref{eq:AR} grow linearly with $|\mu|\tan\beta$ and are thus important for large values of $|\mu| \tan\beta$. They are also proportional to 
the bino mass $|m_{\tilde B}|$, which in mini-split SUSY is much smaller than the slepton mass, roughly by a loop factor. Effectively, the above contribution is thus of two loop size, compared to the case where all mass parameters are at the same scale (as in TeV scale SUSY).

Wino loops can only contribute to $A^L$ and are necessarily proportional to the muon mass.
Compared to the bino contributions \eqref{eq:AL}, \eqref{eq:AR} they arise at linear order in mass insertion, ${\mathcal O}(\delta^L)$, are not  suppressed by a small gaugino mass and are typically dominant for small $|\mu|$.
The general form of the wino contributions to leading order in the mass insertion approximation reads
\begin{equation} \label{eq:AL_wino}
 A_{\mu e}^{L,\tilde W} = \frac{\alpha_2}{4 \pi} \frac{1}{m_{\tilde \ell}^2}  ~(\delta_{\mu e}^L)~ \left[ -\frac{1}{8} g_1(x_W) + g_2(x_W,x_\mu) + \frac{\mu m_{\tilde W}}{m_{\tilde \ell}^2} ~t_\beta~ g_3(x_W,x_\mu) \right] ~,
\end{equation}
with the mass ratios $x_W = |m_{\tilde W}|^2/m_{\tilde \ell}^2$ and $x_\mu = |\mu|^2/m_{\tilde \ell}^2$. Explicit expressions for the loop functions $g_i$ can be found in Appendix~\ref{app:loopfunctions}.
The first term in the parenthesis comes from a pure wino loop. The second and third term involve wino-higgsino mixing and the corresponding diagrams are shown in the middle and right panels of Fig~\ref{fig:muegamma}.
For $|m_{\tilde W}| \ll m_{\tilde \ell}$, the loop function $g_1$ in the first term reduces to $g_1(x) \xrightarrow{x \to 0} 1$.
The loop functions $g_2$ and $g_3$ depend both on the wino and the higgsino mass. For heavy higgsinos, {\it i.e.} in the limit $|m_{\tilde W}| \ll |\mu| \simeq m_{\tilde \ell}$, we find to leading order
\begin{equation}
 g_2(x,y) \xrightarrow{x \to 0} \frac{-11 -7y}{4(1-y)^3} - \frac{(2+6y+y^2)}{2(1-y)^4} \log y ~,~~~ 
 g_3(x,y) \xrightarrow{x \to 0} \frac{1}{y} \log x ~.
\end{equation}
For light higgsinos instead, {\it i.e.} in the limit $|m_{\tilde W}| \simeq |\mu| \ll m_{\tilde \ell}$, we get
\begin{eqnarray}
 g_2(x,y) \xrightarrow{x,y \to 0} \frac{x \log x}{y-x} + \frac{y \log y}{x-y} ~, ~~~
 g_3(x,y) \xrightarrow{x,y \to 0} \frac{\log x }{y-x} + \frac{\log y }{x-y} ~.
\end{eqnarray}
Large logs appear in these expressions, that can enhance the wino loops compared to the bino loops.
For small values of $|\mu| \sim |m_{\tilde B}|, |m_{\tilde W}|$, the wino loops generically dominate. For large values of $|\mu| \sim m_{\tilde \ell}$, bino and wino loops are typically comparable.

\begin{figure}[tb] \centering
\includegraphics[width=0.46\textwidth]{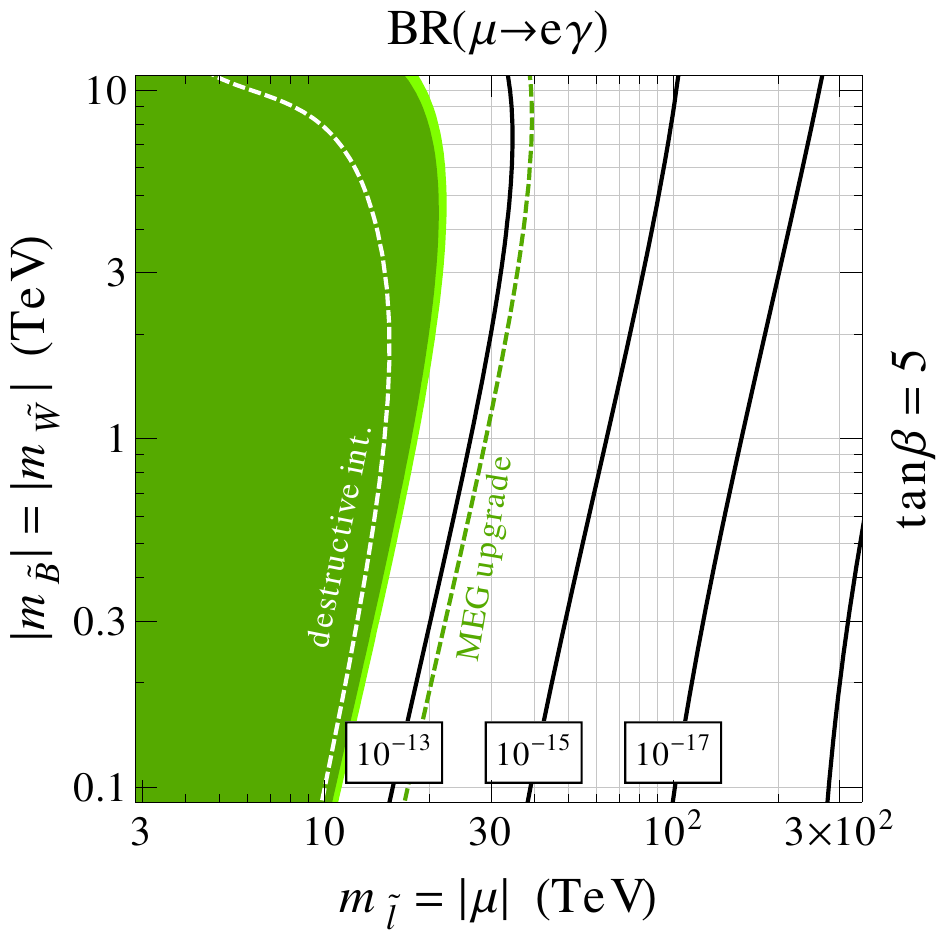} ~~~~~~~~ \includegraphics[width=0.46\textwidth]{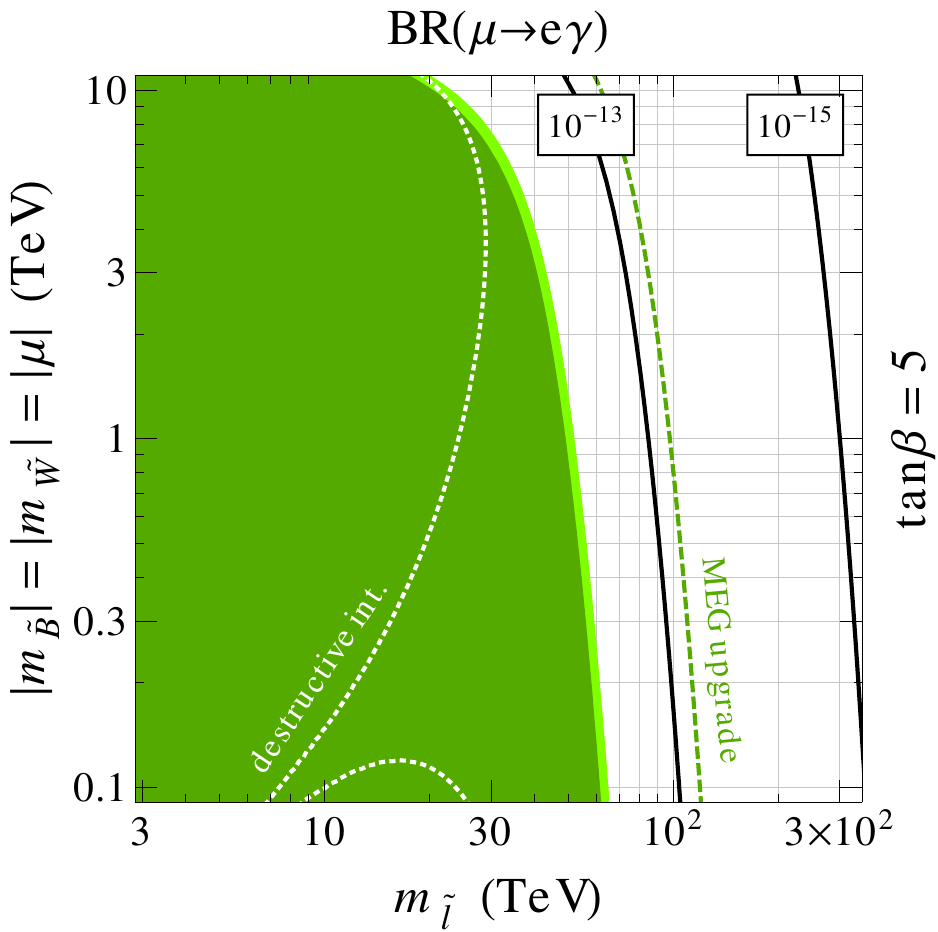} \\[12pt]
\includegraphics[width=0.46\textwidth]{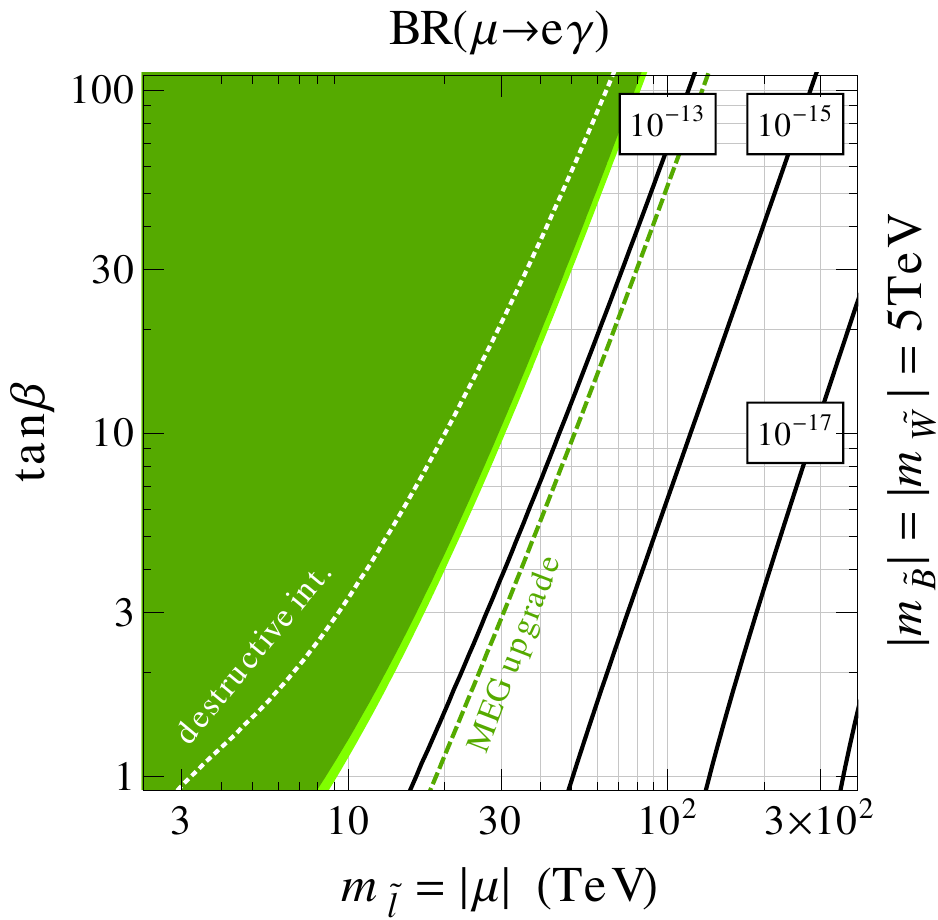} ~~~~~~~~ \includegraphics[width=0.46\textwidth]{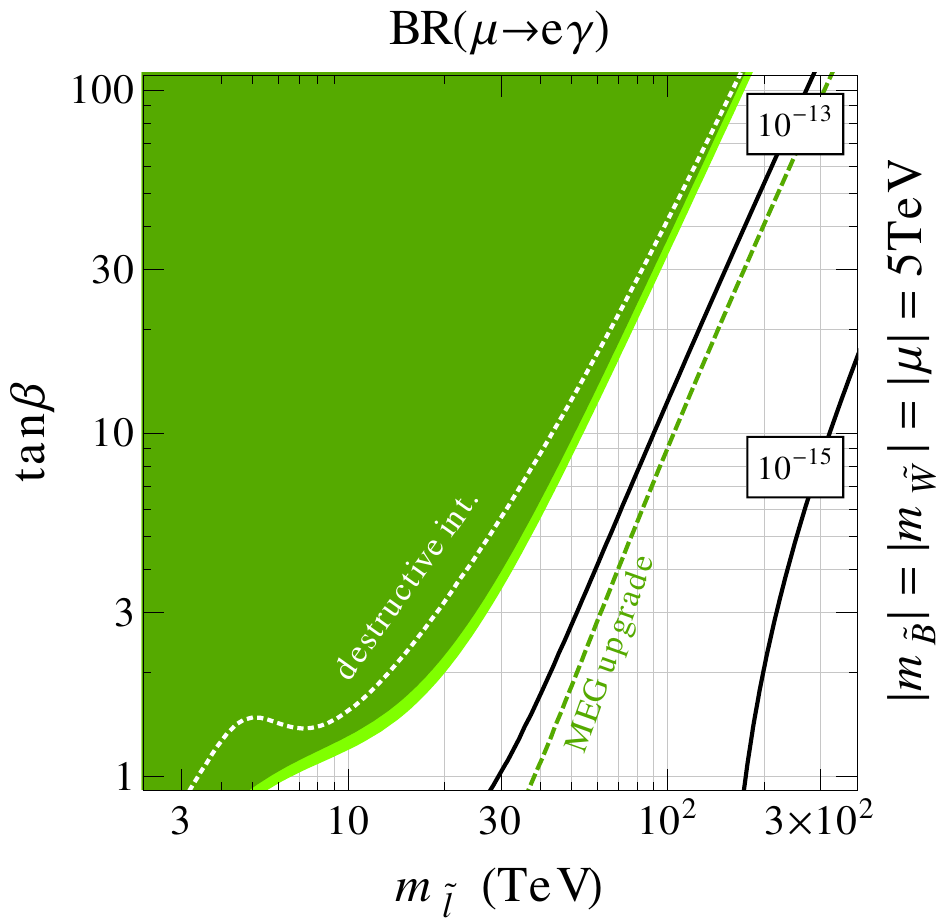}
\caption{\small
Bounds from $\mu \to e \gamma$ in the $m_{\tilde \ell}$ vs. $|m_{\tilde B}| = |m_{\tilde W}|$ plane (top row) and the $m_{\tilde \ell}$ vs. $\tan\beta$ plane (bottom row). The higgsino mass is at the scale of the slepton masses (left column) or at the scale of the gaugino masses (right column). All relevant mass insertions are set to $|\delta_{ij}^L| = |\delta_{ij}^R| = 0.3$. The dark (light) shaded regions are excluded at the 95\% (90\%) C.L. by the current measurement assuming constructive interference between the respective dominant NP amplitudes. The white dotted lines show the case of destructive interference. The dashed lines show the sensitivity of the proposed MEG upgrade.}
\label{fig:mu2egamma}
\end{figure}

The plots in Fig.~\ref{fig:mu2egamma} show the regions of parameter space that are probed by current and future measurements of the $\mu \to e \gamma$ branching ratio. Bounds are shown in the $m_{\tilde \ell}$ vs. $|m_{\tilde B}| = |m_{\tilde W}|$ plane with a fixed $\tan\beta = 5$ (top row) and the $m_{\tilde \ell}$ vs. $\tan\beta$ plane with fixed gaugino masses $|m_{\tilde B}| = |m_{\tilde W}| = 5$~TeV (bottom row). The Higgsino mass is at the scale of the slepton masses (left column) or at the scale of the gaugino masses (right column). All relevant mass insertions are set to $|\delta_{ij}^L| = |\delta_{ij}^R| = 0.3$. The dark (light) shaded regions are excluded at the 95\% (90\%) C.L. by the current measurement. We choose the signs of $\mu m_{\tilde B}$ and $\mu m_{\tilde W}$ such that the different contributions in~(\ref{eq:AL}) and~(\ref{eq:AL_wino}) interfere constructively. The white dotted lines show the case of destructive interference between the dominant terms in each case. The dashed lines show 
the sensitivity of the proposed MEG upgrade.

As expected, the bounds become stronger for larger $\tan\beta$. Current bounds on the slepton masses reach roughly from ${\mathcal O}$(10) TeV (small $\tan\beta$) up to ${\mathcal O}$(100) TeV (large $\tan\beta$). The proposed MEG upgrade can improve the bounds by another factor of $\sim 2$. As indicated by the white dotted lines, such bounds are to some extent model dependent. There exist fine-tuned regions of parameter space where cancellations among the various contributions occur and the bounds can be relaxed considerably.

\subsection{\texorpdfstring{\boldmath $\mu \to e$}{mu --> e} Conversion in Nuclei} \label{sec:mu2e}

The current most stringent experimental bound on the $\mu \to e$ conversion rate was obtained by the SINDRUM II collaboration using gold nuclei~\cite{Bertl:2006up},
\begin{equation} \label{eq:mu2e_exp}
 {\rm BR}^\text{Au}_{\mu \to e} < 7 \times 10^{-13} ~~~@~90\%~ \textnormal{C.L.}~.
\end{equation}
Proposed experiments aim at sensitivities of $\text{BR}^\text{Al}_{\mu \to e} < 10^{-16}$ using Al~\cite{Hewett:2012ns,Abrams:2012er}. In the long term, sensitivities down to $\text{BR}^\text{Al}_{\mu \to e} < 10^{-18}$ might be possible. This corresponds to an improvement by almost 4 to 6 orders of magnitude compared to the result in~(\ref{eq:mu2e_exp}).

The branching ratio of $\mu \to e$ conversion in nuclei can be written as~\cite{Kitano:2002mt}
\begin{eqnarray}
 {\rm BR}^N_{\mu \to e} \times \omega_{\rm cap.}^N &=& \Big| \frac{1}{4} A_{\mu e}^L D + 2(2C_{LV}^u + C_{LV}^d) V^{(p)} +2(C_{LV}^u + 2C_{LV}^d) V^{(n)}\Big|^2 \nonumber \\
&& + \Big| \frac{1}{4} A_{\mu e}^R D + 2(2C_{RV}^u + C_{RV}^d) V^{(p)} +2(C_{RV}^u + 2C_{RV}^d) V^{(n)}\Big|^2~.
\end{eqnarray}
Here, $\omega_{\rm cap.}^N$ is the muon capture rate of the nucleus $N$, and $D$, $V^{(p)}$, and $V^{(n)}$ are nucleus dependent overlap integrals~\cite{Kitano:2002mt}.
The coefficients $A_{\mu e}^{L,R}$ were already introduced in the discussion of the $\mu \to e \gamma$ decay, Eq.  \eqref{eq:Heff_muegamma}. The remaining coefficients are the Wilson coefficients multiplying the dimension six operators in the effective Hamiltonian
\begin{equation}
 \mathcal{H_{\rm eff}} = C_{LV}^q (\bar e \gamma_\nu P_L \mu)(\bar q \gamma^\nu q) + C_{RV}^q (\bar e \gamma_\nu P_R \mu)(\bar q \gamma^\nu q) ~,
\end{equation}
which describes the effective interaction of $\mu$ and $e$ with a vector quark current.
Interactions with axial-vector, pseudo-scalar and tensor quark currents do not contribute to a coherent conversion process and can be neglected.
Interactions with scalar quark currents are suppressed by small lepton and quark masses in the mini-split SUSY framework and are therefore also negligible.

\begin{figure}[tb] \centering
\includegraphics[width=0.29\textwidth]{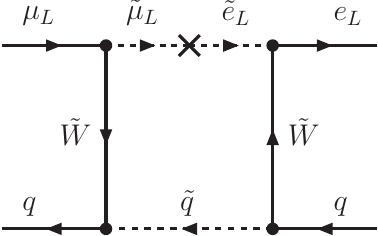} ~~~~~~~ \includegraphics[width=0.29\textwidth]{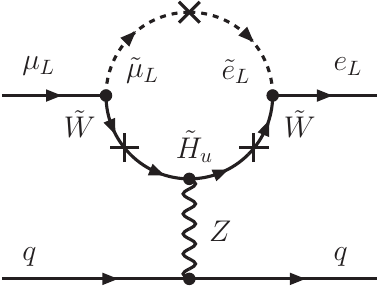} ~~~~~~~ \includegraphics[width=0.29\textwidth]{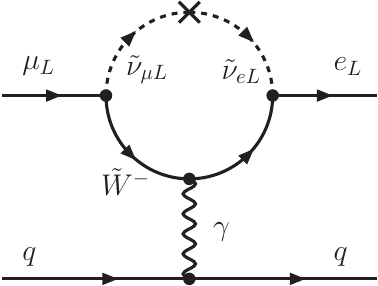}
\caption{\small Example contributions to the amplitude of $\mu \to e$ transition in nuclei. Left: wino box contribution. Middle: $Z$ penguin contribution. Right: photon penguin contribution.}
\label{fig:mu2e_diagrams}
\end{figure}

There are various SUSY contributions to $\mu \to e$ conversion. We already discussed the contributions to the $A^{L,R}_{\mu e}$
in Sec.~\ref{sec:muegamma}. These dipole contributions are by far the dominant terms in $\mu \to e$ conversion for the frequently studied framework of TeV scale SUSY with sizable $\tan\beta$. In mini-split SUSY, on the other hand, the 4 fermion operators cannot be neglected.
Contributions to the 4 fermion operators can come from boxes, photon penguins, and $Z$ penguins. Example diagrams are shown in Fig.~\ref{fig:mu2e_diagrams}.

The largest contribution to the box diagrams comes from wino loops, shown in Fig.~\ref{fig:mu2e_diagrams} (left). Since winos only interact with the left handed (s)fermions, the wino box diagrams do not contribute to $C_{RV}^q$. For the contributions to $C_{LV}^q$ we find in the limit $|m_{\tilde W}| \ll m_{\tilde \ell}, m_{\tilde q}$ and to first order in the mass insertions
\begin{equation}
 5 C_{LV}^{u, \text{box}} = C_{LV}^{d, \text{box}} = \frac{\alpha_2^2}{m_{\tilde q}^2} ~(\delta^L_{\mu e})~ \frac{5}{4} f\left({m_{\tilde \ell}^2}/{m_{\tilde q}^2}\right) ~, 
\end{equation}
with the loop function 
\begin{equation}
 f(x) = \frac{1}{8(1-x)} + \frac{x\log x}{8(1-x)^2} ~,{\rm{~so~that}}~~ f(1) = \frac{1}{16}~,~~ f(0) = \frac{1}{8} ~.
\end{equation}
The wino boxes decouple if either the squark mass $m_{\tilde q}$ or slepton mass $m_{\tilde \ell}$ become large. They do not contain large logs, are largely independent of the gaugino masses and also independent of the $\mu$ parameter. 

The photon penguins are also dominated by wino loops, see Fig.~\ref{fig:mu2e_diagrams} (right), which  generate only  the left-handed coefficients $C_{LV}^q$,
\begin{equation}
 -\frac{1}{2} C_{LV}^{u, \gamma} = C_{LV}^{d, \gamma} = \frac{\alpha_\text{em} \alpha_2}{m_{\tilde \ell}^2} ~(\delta^L_{\mu e})~  \left[ \frac{1}{4} + \frac{1}{9} \log\left(\frac{|m_{\tilde W}|^2}{m_{\tilde \ell}^2}\right) \right] ~.
\end{equation}
The ratio between the photon penguin contributions to $C_{LV}^u$ and $C_{LV}^d$ is set by the quark charges.
Note that the photon penguin is enhanced by a large logarithm, $\log(|m_{\tilde W}|^2/m_{\tilde \ell}^2)$, which arises from diagrams where the photon couples to the light charged wino (as in the right diagram of Fig.~\ref{fig:mu2e_diagrams}).

Finally, $Z$ penguins arise dominantly from diagrams that involve higgsino-wino mixing. The general form of the $Z$ penguin contributions to $C_{LV}^q$ reads
\begin{eqnarray}
 && C_{LV}^{d, Z} = \frac{\alpha_2^2}{m_{\tilde \ell}^2} ~(\delta^L_{\mu e})~ \frac{1}{16}\left( 1 - \frac{4}{3} s_W^2 \right) \left[ c_\beta^2 f_1(x_W,x_\mu) + s_\beta^2 f_2(x_W,x_\mu) + \frac{\mu m_{\tilde W}}{m_{\tilde \ell}^2} s_\beta c_\beta f_3(x_W,x_\mu) \right] ~, \nonumber \\
 && C_{LV}^{u, Z}\left( 1 - \frac{4}{3} s_W^2 \right) = - \left( 1 - \frac{8}{3} s_W^2 \right) C_{LV}^{d, Z} ~.
\end{eqnarray}
As before, $x_W = |m_{\tilde W}|^2/m_{\tilde \ell}^2$ and $x_\mu = |\mu|^2/m_{\tilde \ell}^2$. The full form of the loop functions $f_i$ is given in Appendix~\ref{app:loopfunctions}.
In the limit of light winos and heavy higgsinos, we find for the loop functions
\begin{equation}
 f_1(x,y),~ 3f_2(x,y) \xrightarrow{x \to 0} \frac{3y -9}{2(1-y)^2} - \frac{3 \log y}{(1-y)^3} ~,~~~ 
 f_3(x,y) \xrightarrow{x \to 0} -\frac{12(1+y)}{y(1-y)^2} - \frac{24 \log y}{(1-y)^3} ~.
\end{equation}
In the other relevant limit, namely light winos and light higgsinos we find
\begin{equation}
 f_1(x,y),~ 3f_2(x,y) \xrightarrow{x,y \to 0} -3 \log y ~,~~~ 
 f_3(x,y) \xrightarrow{x,y \to 0} -24 \log y ~.
\end{equation}
Note the appearance of large logs in the case of a small higgsino mass.

\begin{figure}[tb] \centering
\includegraphics[width=0.46\textwidth]{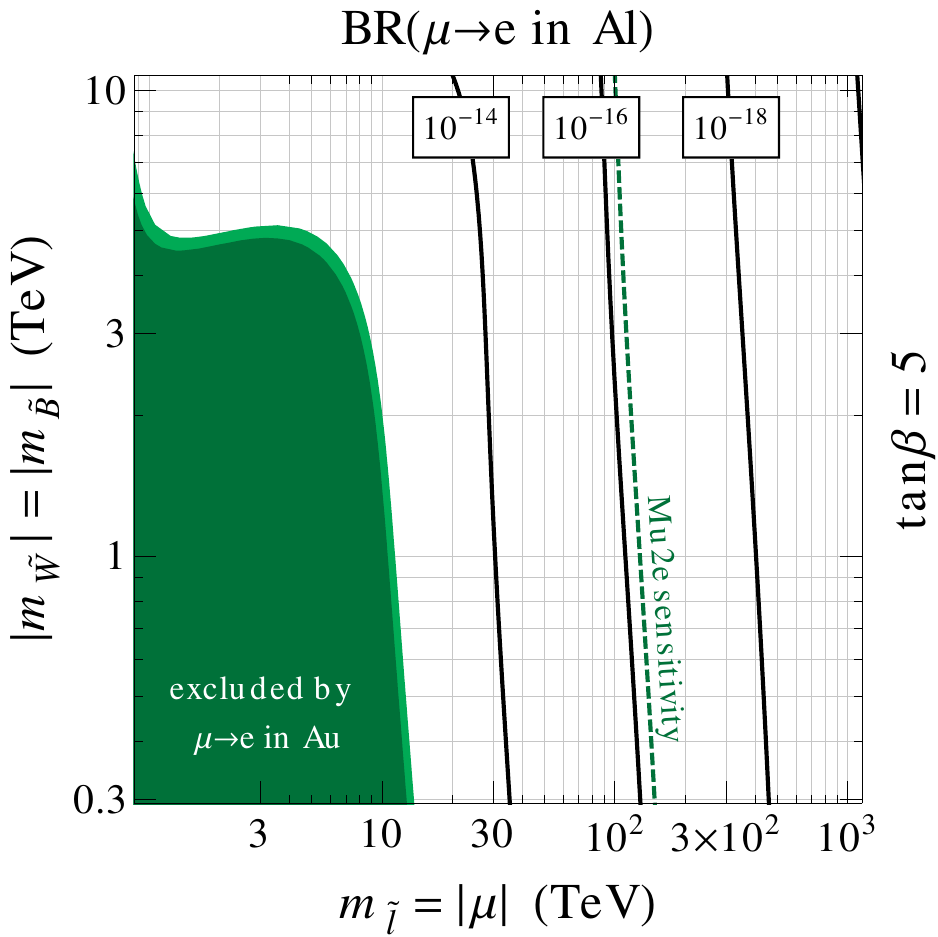} ~~~~~~~~ \includegraphics[width=0.46\textwidth]{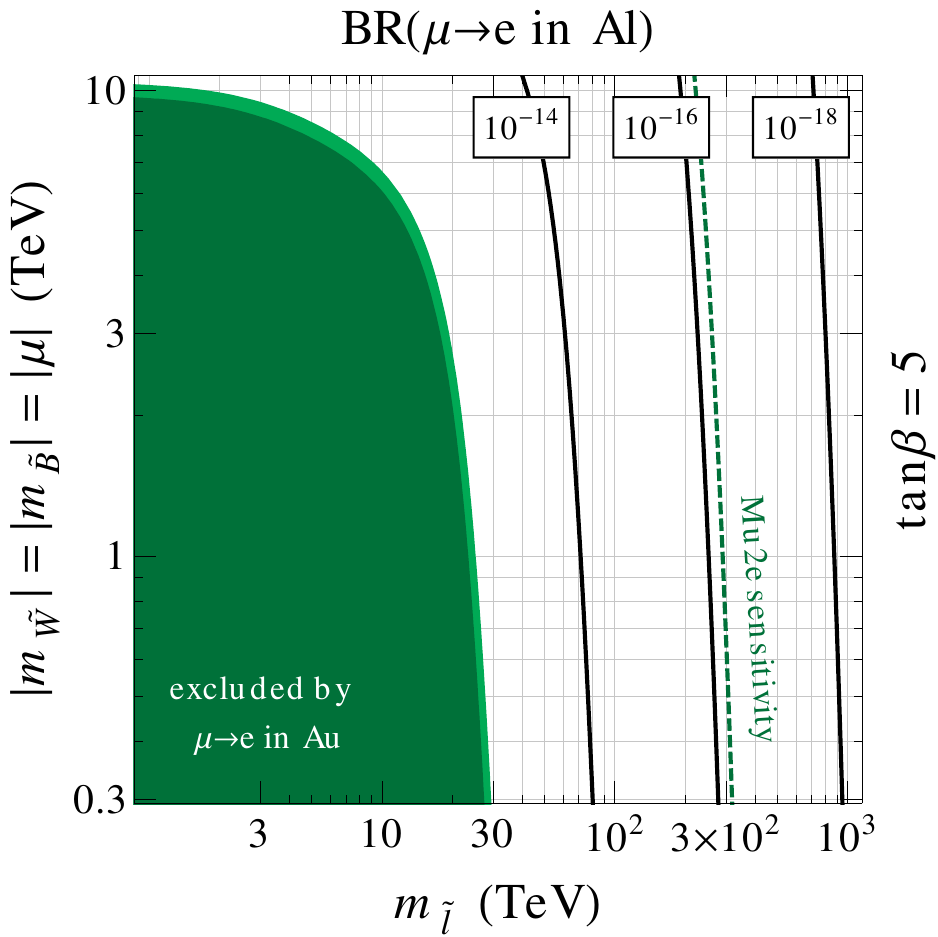} \\[12pt]
\includegraphics[width=0.46\textwidth]{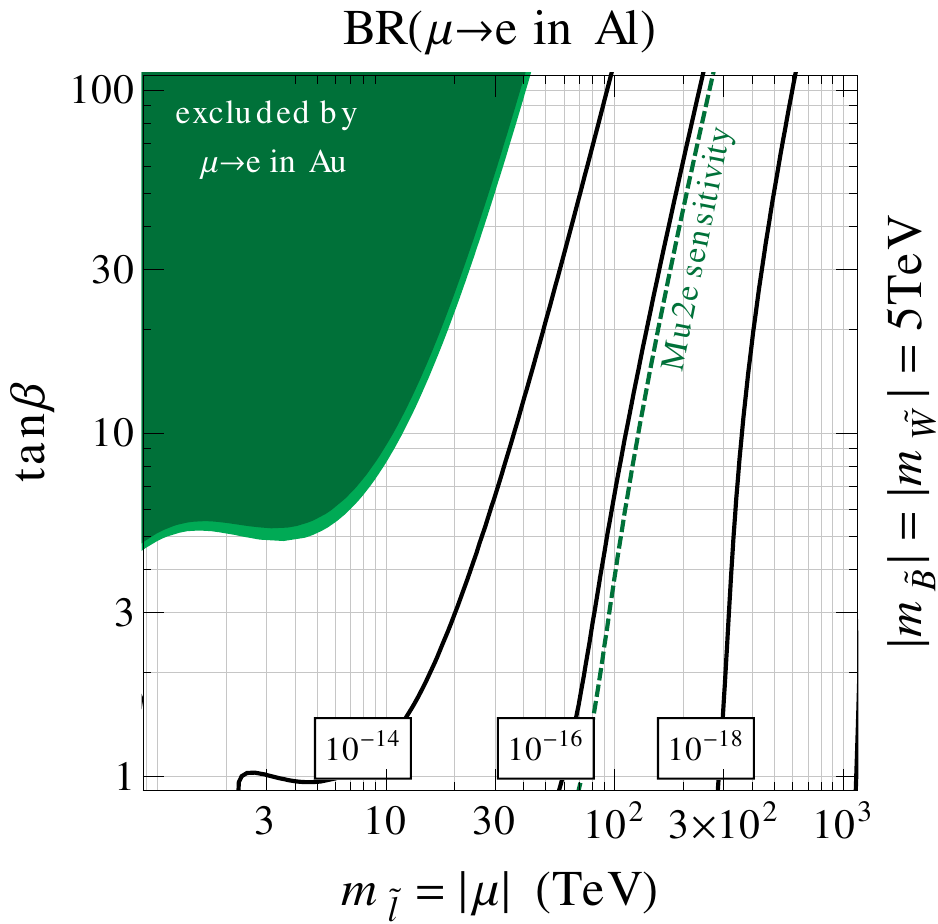} ~~~~~~~~ \includegraphics[width=0.46\textwidth]{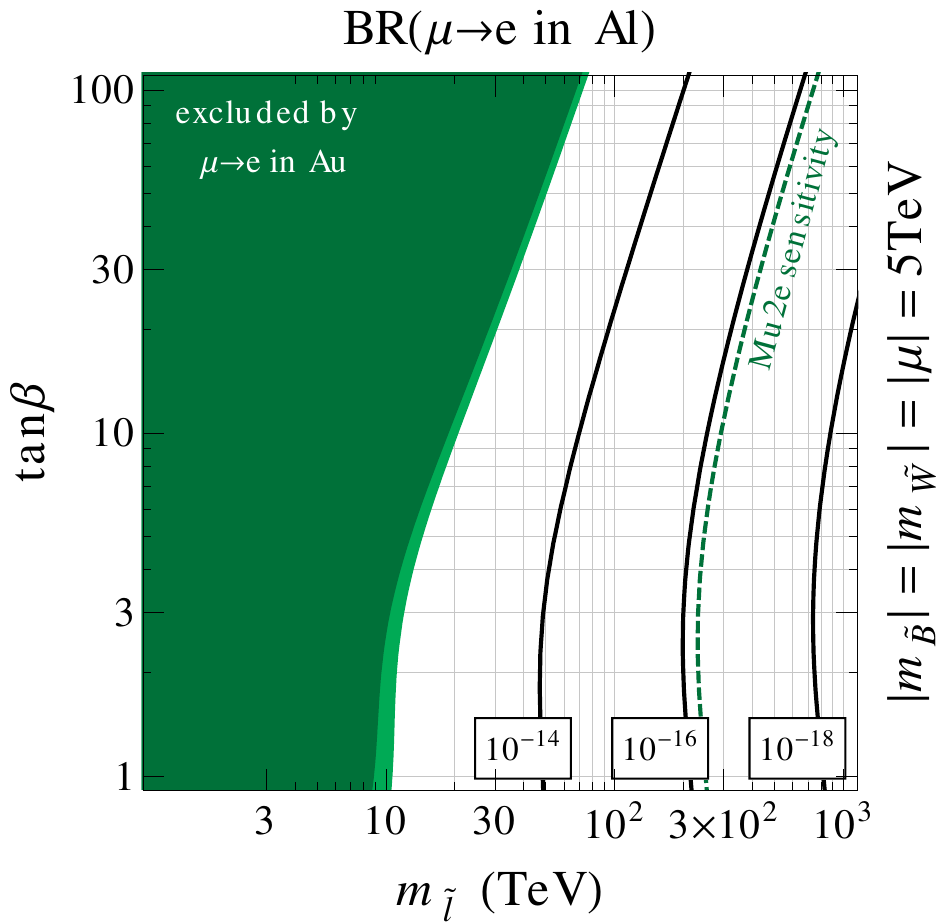}
\caption{\small
Predicted $\mu \to e$ conversion rates in Al in the $m_{\tilde \ell}$ vs. $|m_{\tilde B}| = |m_{\tilde W}|$ plane (top row) and the $m_{\tilde \ell}$ vs. $\tan\beta$ plane (bottom row). The higgsino mass is set either equal to the slepton masses (left column) or to the gaugino masses (right column). All relevant mass insertions are fixed to $|\delta_{ij}^L| = |\delta_{ij}^R| = 0.3$. The dark (light) shaded regions show 95\% (90\%) C.L. exclusions by the current limits on $\mu \to e$ conversion in Au, while the sensitivity of the planned Mu2e experiment is given by the dashed lines.}
\label{fig:mu2e}
\end{figure}

For large values of $\tan\beta$, the contributions from the dipoles are typically dominant. For moderate $\tan\beta$, however, the other contributions can become important as well. For small Higgsino mass $|\mu| \sim |m_{\tilde B}| , |m_{\tilde W}|$ and moderate $\tan\beta$, both $Z$ penguins and photon penguins usually give  the largest contributions due to the $\log(|\mu|^2/m_{\tilde \ell}^2)$ and $\log(|m_{\tilde W}|^2/m_{\tilde \ell}^2)$ enhancement.  For a large $\mu$ parameter (and moderate $\tan\beta$), only the photon penguins are enhanced and generically give the dominant SUSY contributions to the $\mu \to e$ conversion. Box contributions are typically only relevant in the regime where $|m_{\tilde W}| \sim m_{\tilde \ell}, m_{\tilde q}$ which is contrary to the spirit of the mini-split SUSY setup.

Current and expected bounds on the mini-split SUSY parameter space from $\mu \to e$ conversion are shown in Fig.~\ref{fig:mu2e}. The dark (light) shaded regions are excluded at the 95\% (90\%) C.L. by the current limits on the $\mu \to e$ conversion rate in Au. The black solid lines show the predicted rates for the $\mu \to e$ conversion rate in Al, with the expected sensitivity of the Mu2e experiment indicated by a dashed line. The bounds are either shown in the $m_{\tilde \ell}$ vs. $|m_{\tilde B}| = |m_{\tilde W}|$ plane with a fixed $\tan\beta = 5$ (top row) or in the $m_{\tilde \ell}$ vs. $\tan\beta$ plane, where gaugino masses are fixed to $|m_{\tilde B}| = |m_{\tilde W}| = 5$~TeV (bottom row). We also show a two-fold choice for the value of the $\mu$ parameter. It is either set to be equal to the slepton masses (left column) or to the gaugino masses (right column). In all panels the signs of the gaugino masses and the mass insertions (taken to be $|\delta_{ij}^L| = |\delta_{ij}^R| = 0.3$)  are chosen 
such that the dominant contributions interfere constructively for large slepton masses.  

From Fig.~\ref{fig:mu2e} one sees that for large $\tan\beta$ the constraints become stronger with increasing $\tan\beta$ due to dipole dominance. For small $\tan\beta$, however, $Z$ or photon penguins dominate and the bounds become approximately independent of $\tan\beta$. For moderate $\tan\beta$, current bounds on the slepton masses are around ${\mathcal O}$(10 TeV) for large $|\mu|$ and around ${\mathcal O}$(30 TeV) for small $|\mu|$. The sensitivity of the Mu2e experiment will allow to improve these bounds by at least one order of magnitude and to probe slepton masses generically at a scale of ${\mathcal O}$(300 TeV) and above, barring accidental cancellations.

\subsection{The \texorpdfstring{\boldmath $\mu \to 3 e$}{mu --> 3 e} Decay} \label{sec:mu3e}

\begin{figure}[tb] \centering
\includegraphics[width=0.46\textwidth]{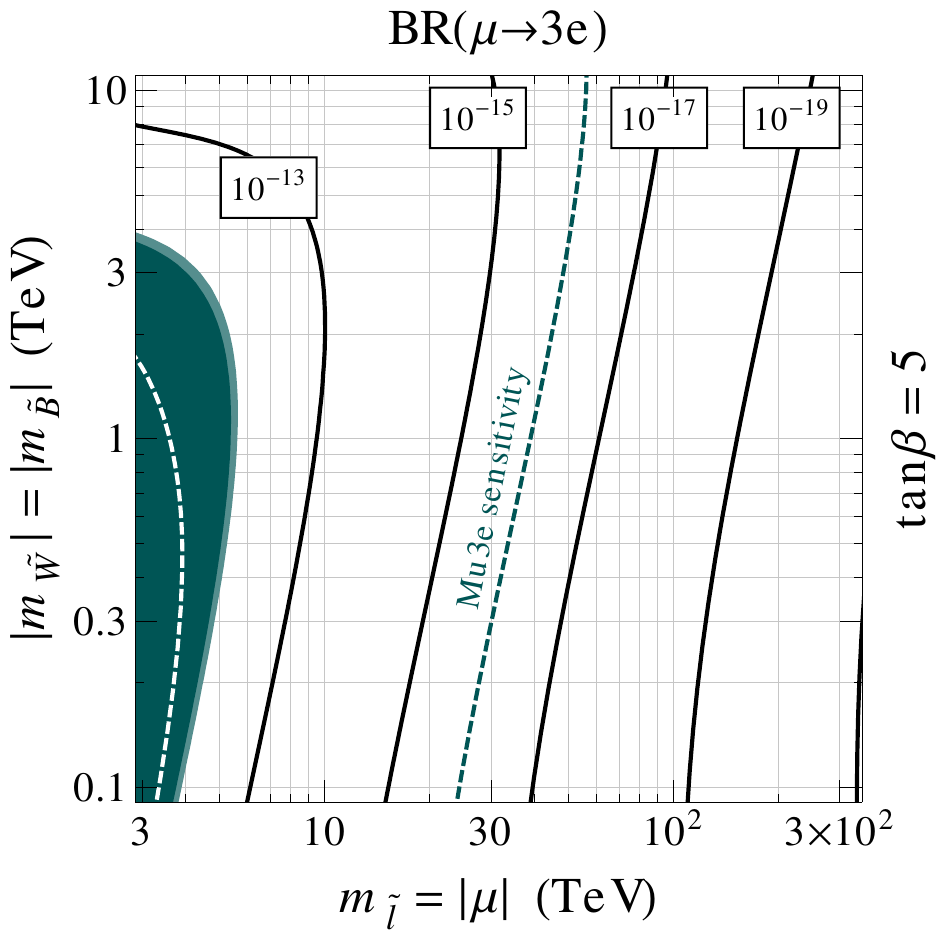} ~~~~~~~~ \includegraphics[width=0.46\textwidth]{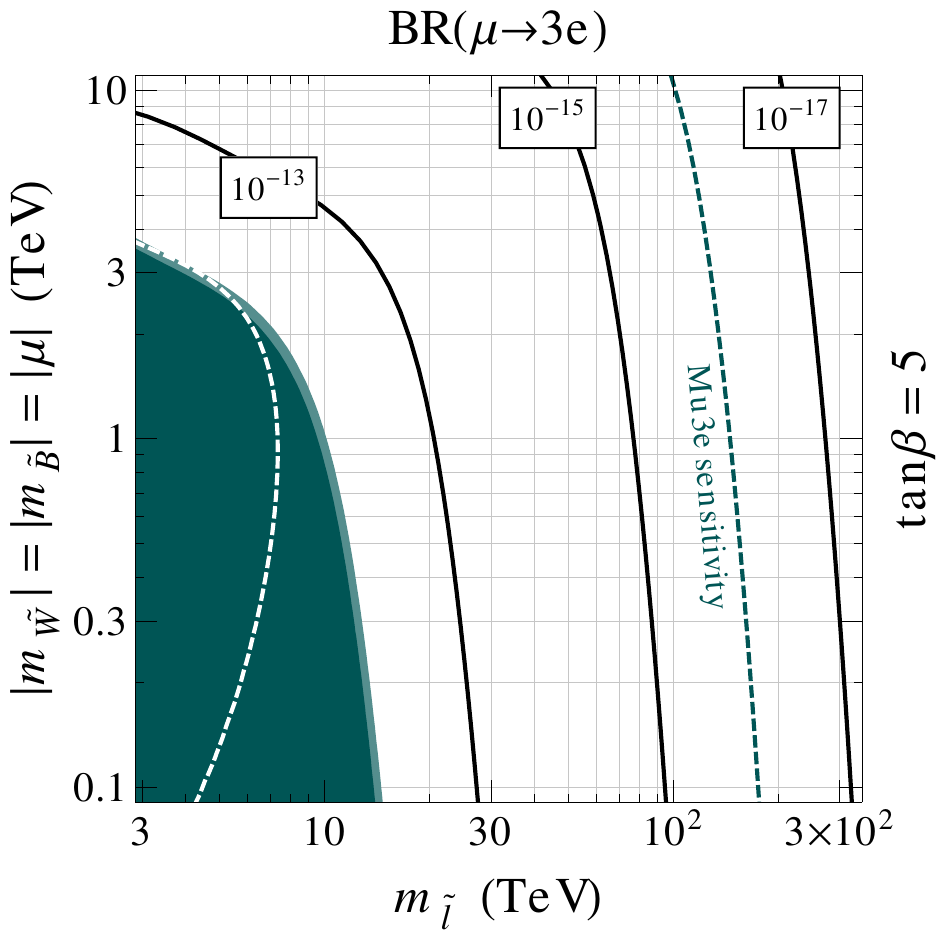} \\[12pt]
\includegraphics[width=0.46\textwidth]{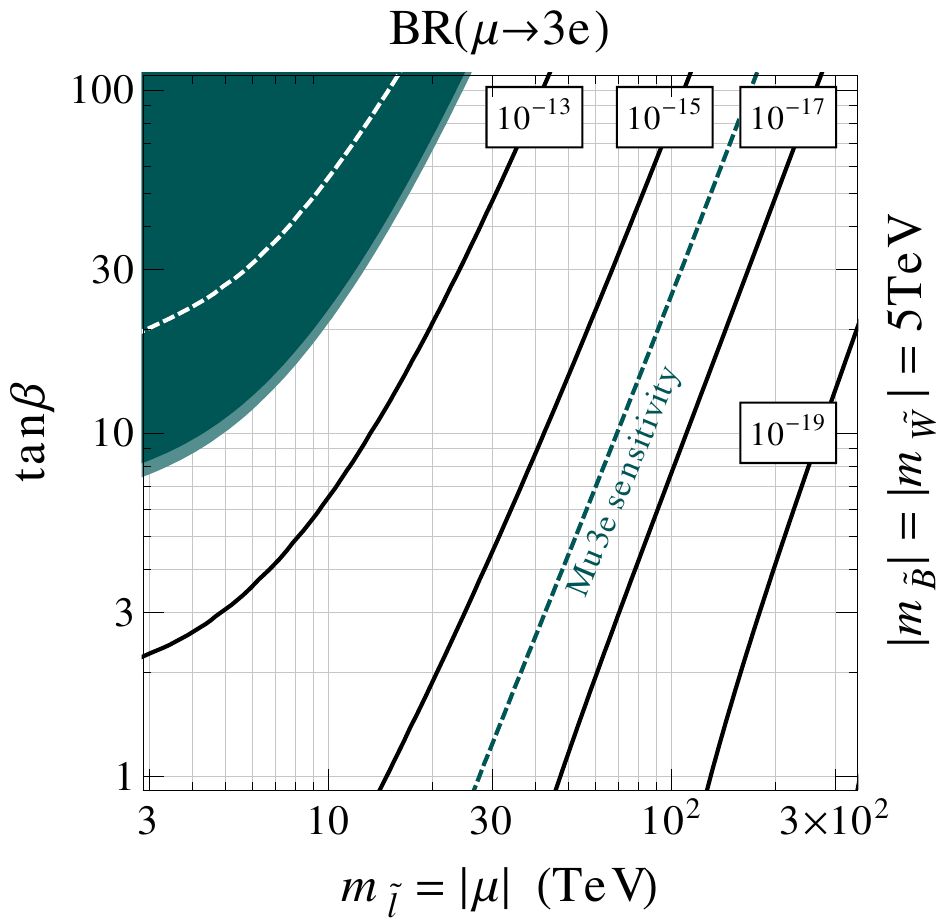} ~~~~~~~~ \includegraphics[width=0.46\textwidth]{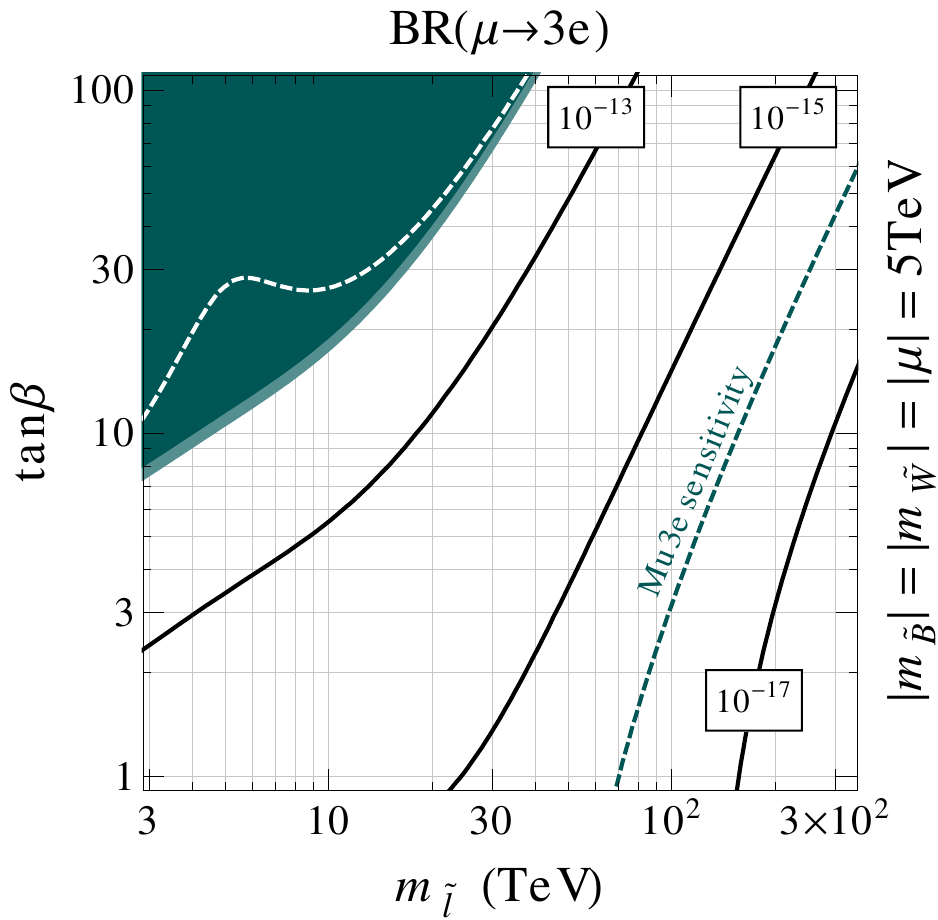}
\caption{\small
Bounds from the $\mu \to 3 e$ decay in the $m_{\tilde \ell}$ vs. $|m_{\tilde B}| = |m_{\tilde W}|$ plane (top row) and the $m_{\tilde \ell}$ vs. $\tan\beta$ plane (bottom row). The higgsino mass is either equal to slepton masses (left column) or to the gaugino masses (right column). All relevant mass insertions are set to $|\delta_{ij}^L| = |\delta_{ij}^R| = 0.3$. The dark (light) shaded regions are excluded at the 95\% (90\%) C.L. by the current measurement assuming constructive interference between the respective dominant NP amplitudes. The white dotted lines show the case of destructive interference. The dashed lines show the sensitivity of the proposed Mu3e experiment.}
\label{fig:mu3e}
\end{figure}

In the frequently studied case of TeV scale SUSY, it is well known that even for moderate and low $\tan\beta$ the $\mu \to 3 e$ decay rate is dominated by the contributions from the $\mu \to e \gamma$ dipole operator~\cite{Hisano:1995cp,Arganda:2005ji}. The dipole dominance is due to the appearance of a large $\log(m_\mu^2/m_e^2)$ in the corresponding phase space integration. In the mini-split SUSY framework, thus dipole dominance typically remains a good approximation. It gives
\begin{equation}
 \frac{{\rm BR}(\mu \to 3e)}{{\rm BR}(\mu \to e \gamma)} \simeq \frac{\alpha_{\rm em}}{3 \pi} \left( \log\left(\frac{m_\mu^2}{ m_e^2}\right) -\frac{11}{4} \right) \simeq 6 \times 10^{-3}.
\end{equation}
The current bound on the branching ratio of the $\mu \to 3 e$ decay~\cite{Bellgardt:1987du}
\begin{equation}
 {\rm BR}(\mu \to 3e) < 1.0 \times 10^{-12} ~~~@~90\%~ \textnormal{C.L.}~,
\end{equation}
leads to much weaker constraints than the direct constraint from the $\mu \to e \gamma$ branching ratio~(\ref{eq:muegamma_exp}). The proposed Mu3e experiment~\cite{Blondel:2013ia} aims at an ultimate sensitivity of BR$(\mu \to 3 e) \lesssim 10^{-16}$. Assuming dipole dominance, this would lead to constraints slightly better than to those obtained from the MEG update~\cite{Baldini:2013ke}. 

The bounds on the mini split SUSY parameter space implied by $\mu \to 3 e$ are shown in the plots in Fig.~\ref{fig:mu3e}, in complete analogy to the $\mu \to e \gamma$ plots in Fig.~\ref{fig:mu2egamma}. We explicitly checked that contributions from boxes, photon penguins, and $Z$ penguins are generically small. Nonetheless, we include them in our numerics using the general expressions for the branching ratio given in~\cite{Hisano:1995cp,Arganda:2005ji}. As anticipated, current bounds from $\mu \to 3 e$ are significantly weaker compared to the $\mu \to e \gamma$ bounds. Only for very large values of $\tan\beta$ can $\mu \to 3 e$ probe slepton masses beyond $10$~TeV. The sensitivity of the proposed Mu3e experiment, on the other hand, will allow to probe scales of 100 TeV and beyond even for moderate $\tan\beta \sim 5$.

\section{Implications for Models of Fermion Masses}\label{sec:fermion-masses}

So far we considered mini-split SUSY in a ``generic'' setting, i.e., large ${\mathcal O}(1)$ flavor violation at a universal squark and slepton mass scale. 
It is interesting though to put our results also in context of models that explain the SM flavor structure. We consider two possibilities -- models with horizontal symmetries 
and models with radiative mass generation.
In the models with horizontal symmetries the SM fermion masses and mixings suggest textures for flavor violating entries in the squark and slepton mass matrices. 
In models with radiative fermion mass generation, on the other hand, the creation of first generation fermion masses through scalar loops  requires a minimal amount of flavor violation, which can then be tested with low energy probes.

\subsection{Textures: Anarchy versus Hierarchy.}

One of the most popular ways to explain the flavor structure of the SM is to invoke a flavor texture for the Yukawa matrices. The simplest such models include a set of flavor dependent spurions  $\epsilon_A$ associated with the breaking of chiral symmetry 
\begin{equation}
y_u^{ij}\sim \epsilon_Q^i \epsilon^j_U ~,\qquad y_d^{ij}\sim \epsilon^i_Q \epsilon^j_D ~,\qquad y_l^{ij}\sim \epsilon^i_L \epsilon^j_E ~.
\label{eq:texture-fermions}
\end{equation}
Such textures arise in many models, including models with horizontal symmetries~\cite{Leurer:1992wg,Leurer:1993gy}, Froggatt-Nielsen models~\cite{Froggatt:1978nt}, models with extra dimensions, either flat~\cite{Kaplan:2001ga} or warped~\cite{Gherghetta:2000qt, Agashe:2004cp, Grossman:1999ra}, Nelson-Strassler models~\cite{Nelson:2000sn} (generation of hierarchy by large anomalous dimensions), etc. 
It is useful to review here a choice for $\epsilon_A$ that produces  qualitative agreement with the known fermion masses and mixings. As a concrete and well motivated benchmark we also impose $SU(5)$ relations which gives (see e.g.~\cite{Nelson:2000sn})
\begin{equation}
\label{eq:epsilons}
\epsilon_{10}\equiv\epsilon_Q=\epsilon_U=\epsilon_E\sim
\begin{pmatrix} 0.003 \cr 0.04 \cr 1 \end{pmatrix}
\qquad \mbox{and} \qquad 
\epsilon_{\bar 5}\equiv \epsilon_D=\epsilon_L\sim
\begin{pmatrix} 0.004 \cr 0.025 \cr 0.025 \end{pmatrix} ~.
\end{equation}
The hierarchy in spurions for fields that are in the {\boldmath $10$} representation of $SU(5)$ ($Q$, $U$, and $E$) is more pronounced than for fields in the {\boldmath $\bar 5$} representation ($D$ and $L$), to account for the larger hierarchy among up-type quarks. For simplicity, we took $\tan \beta$ of order one, though this choice will not affect the discussion here. Small variations around the numbers in Eq.~(\ref{eq:epsilons}) are certainly possible, since ${\mathcal O}(1)$ numbers can always compensate for a mild change in the $\epsilon_A$'s.

Given a texture for the fermion mass matrix, it is a model dependent question what the texture of the superpartner masses may be. In order to demonstrate this model dependence, we consider a few possible textures for sfermion mass matrices and give an example of a UV model for each. We also keep track of how the various low energy probes are affected by our choice of the sfermion texture.

\paragraph{\underline{Anarchic sfermion masses}} - This is the ansatz we have been considering thus far. 
The texture for scalar masses is
\begin{equation}
\tilde m_{ij}^2\sim \begin{pmatrix} 
1 & 1 & 1 \cr 
1 & 1 & 1 \cr 
1 & 1 & 1
\end{pmatrix} ~,
\end{equation}
for all scalars. Beyond being phenomenologically convenient and interesting, this ansatz can easily come from simple UV models of flavor. This is precisely the type of scenario generations of model builders have worked to avoid within TeV scale SUSY. It will generally arise in a situation where SUSY breaking and the generation of Yukawa matrices happen at very different scales or different locations in an extra dimension. For example, if SUSY breaking is mediated at a scale far above a Frogatt-Nielsen scale, the scalar masses will be maximally mis-aligned with the flavor basis.

In order to have a concrete example, consider a supersymmetric Kaplan-Tait~\cite{Kaplan:2001ga} model with a flat extra dimension (the scale of the extra dimension is not of importance here) and the Higgs fields confined to an orbifold fixed point at~$x_5=0$. Matter fields live in the bulk of the extra dimension and have flavor dependent bulk masses which are not particularly hierarchical. The profiles of the fermion wave functions in the extra dimension are to a good approximation exponentials $\sim e^{M^{(f)}_i x_5}$, where $M^{(f)}_i$ is the bulk mass for $i$-th generation of the fermion $f$. The fermion mass matrix in the 4D theory will be affected by the exponentially small overlaps between the fermion profiles and the Higgs brane. After normalizing the field profiles on the interval 0 to $R$, we find that the spurion $\epsilon_f$ in Eq.~(\ref{eq:texture-fermions}) for the fermion $f$ is
\begin{equation}
\epsilon_f^i\sim e^{-M^{(f)}_iR} ~.
\end{equation}
Fermion hierarchies are thus easy to achieve. A mild hierarchy of less than an order of magnitude, between the smallest and largest bulk mass is needed to produce the numbers in Eq.~(\ref{eq:epsilons}).

In contrast to the Higgs, the hidden sector field which breaks SUSY may live in the bulk of the extra dimension. Let us assume for simplicity that it has flat profile. The scalar soft masses are then proportional to overlap integrals of the various flavors. One quickly finds~(we suppress the representation index $f$) 
\begin{equation}
\delta_{ij} \sim \frac{2 \sqrt{M_i M_j}}{M_i+M_j} ~,
\end{equation}
which is anarchical for the choices of $M_i$'s we require. For TeV scale SUSY, such simple models have been dismissed by models builders in favor of models where SUSY breaking is localized away from the fermions (in~\cite{Kaplan:2001ga} this was done by adding yet another extra dimension and employing gaugino mediation \cite{Chacko:1999mi,Kaplan:1999ac}). The motivation was precisely to avoid the dangerous flavor constraints which are the topic of our work. In the context of PeV scale SUSY we are free to consider the simplest possibility.

\paragraph{\underline{Flavorful sfermion masses}} - 
Here the ansatz for the scalar mass texture is
\begin{equation}
({m_{\tilde f}^2)_{ij}}\sim \epsilon_f^{i*}\epsilon_f^j\,.
\end{equation}
This class of models includes Nelson-Strassler~\cite{Nelson:2000sn} or flavorful supersymmetry~\cite{Nomura:2007ap}, as well as models with warped or flat extra dimensions~\cite{Nomura:2008pt}. The fermion mass hierarchy can be achieved, for example, by large anomalous dimension or in the dual picture by bulk fermion masses in a warped extra dimension. Since we already presented a flat extra dimensional model for the anarchical case, we will continue along this theme also here (similar to~\cite{Nomura:2008pt}). Similarly to the anarchic case, the fermion mass hierarchy is produced naturally using exponential profiles and a Higgs on a brane. However, this time the SUSY breaking field $X$ is localized on the \emph{same} brane as the Higgs. The operator $X^\dagger X Q^\dagger Q$ in the 4D theory will be proportional to the values of the wave functions  at the brane, namely the $\epsilon$'s.

Throughout our analysis we have taken a common scale for all sfermions. In this case the assumption is violated for most scalars. 
{The bounds thus require a replacement $1/m^2 (\delta_{ij})^n \to  1/m_{\rm light}^2  (m^2_{ij}/m_{\rm heavy}^2)^n$.}
Note, however that the left handed sleptons as well as the right handed down squarks retain an anarchical texture even in this ``hierarchical'' case, cf. Eq. \eqref{eq:epsilons}. We thus expect the LFV observables to remain sensitive to high slepton masses. Furthermore, $K-\bar K$ mixing will also be highly constraining, though its sensitivity is somewhat reduced compared to the fully anarchical case because the enhanced LR hadronic matrix element will be replaced by the moderate RR one. 

\paragraph{\underline{Horizontal symmetries and Frogatt-Nielsen}} - In this setup~\cite{Leurer:1992wg,Leurer:1993gy,Froggatt:1978nt} every fermion $f$ in the SM is assigned a charge~$x_f$ under a flavor $U(1)$ symmetry. The symmetry is violated by a spurion  which carries a charge of -1 and whose size is set to be $\lambda\sim 0.2$, of order the Cabibbo angle. The spurion is presumably generated by the vev of a field, but we will not specify its dynamics here.  Our analysis will also not depend on whether 
the horizontal symmetry is a continuous global symmetry, or a discrete remnant of a $U(1)$.

After introducing the spurion every interaction in the SM formally respect the horizontal symmetry. In the Yukawa interactions every fermion $f$ is thus accompanied by the symmetry violating spurion raised to the power $x_f$. This leads to a texture of the form of Eq.~(\ref{eq:texture-fermions}). For concreteness we will adopt the Seiberg-Nir ``master model''~\cite{Leurer:1993gy} as an example of a viable choice for the charges (the model was extended to accommodate leptons in~\cite{BenHamo:1994bq}). 
This model by itself is ruled out in the context of natural SUSY because of excessive FCNC's, and was used as a useful starting point to build more elaborate models such as models with two or more $U(1)$'s. For PeV SUSY the flavor problem is automatically alleviated and we are free to consider the simplest model imaginable.
The $U(1)$ charge assignments, $x^i_f$ are 
\begin{center}
\begin{tabular}{c|ccc}
 ferm./gen. & 1& 2 & 3\\
 \hline
Q & \ 3 \ &\  2 \ & \ 0 \    \\ 
U &  \  3 \  & \  1 \  & \  0 \    \\
D &  \  3 \  & \ 2  & \  2 \    \\
L  & \  3 \ & \  3 \   & \  3 \   \\
E  & \ 5 \  & \ 2 \ & \ 0  \   
\end{tabular}
\end{center}
where the spurion $\lambda$ carries a charge of~-1.
This choice gives $\epsilon^i_f=\lambda^{x^i_f}$ which are within a factor of a few of the values in Eq.~(\ref{eq:epsilons}) for each entry (here $x_f^i$ is the $U(1)$ charge of the field in question with $i$ the generation index). 
It is also perfectly reasonable to consider variations around this model, as ``order one'' couplings can easily compensate for a single power of $\lambda$ here and there. For example, it is easy to pick charges in which the fields $Q$, $U$ and $E$ have similar charges across generations, as do $D$ and $L$ in order to respect $SU(5)$ (as Eq.~(\ref{eq:epsilons}) does). For concreteness, however, we will stay with the master model.

The squark and slepton soft masses come from operators of the form $Q^\dagger Q$, $L^\dagger L$, etc. Flavor diagonal terms are automatically invariant under the $U(1)$, while off diagonal terms are suppressed by 
\begin{equation}
\delta_{ij}\propto \lambda^{|x_f^i-x_f^j|}.
\end{equation}
For the master model choice of charges the squark mass matrices have the following textures
\begin{equation}
m^2_{\tilde q}\sim 
\begin{pmatrix} 
1 & \lambda & \lambda^3 \cr 
\lambda & 1 &  \lambda^2 \cr 
\lambda^3 & \lambda^2 & 1
\end{pmatrix} ~, \qquad
m^2_{\tilde u}\sim 
\begin{pmatrix}
1 & \lambda^2 & \lambda^3 \cr 
\lambda^2 & 1 &  \lambda \cr 
\lambda^3 & \lambda & 1
\end{pmatrix} ~, \qquad
m^2_{\tilde d}\sim 
\begin{pmatrix}
1 & \lambda & \lambda \cr 
\lambda & 1 &  1 \cr 
\lambda & 1 & 1
\end{pmatrix} ~, \label{eq:squark:masses}
\end{equation}
while the slepton mass textures are
\begin{equation}
m^2_{\tilde l}\sim 
\begin{pmatrix} 
1 & 1 & 1 \cr 
1 & 1 & 1 \cr 
1 & 1 & 1
\end{pmatrix}~,  \qquad
m^2_{\tilde e}\sim 
\begin{pmatrix} 
1 & \lambda^2 & \lambda^5 \cr 
\lambda^2 & 1 &  \lambda^3 \cr 
\lambda^5 & \lambda^3 & 1
\end{pmatrix}~.\label{eq:slepton:masses}
\end{equation}
In Tab.~\ref{Tab:lambda-supression}, we show which mass insertions control the dominant contribution to the various observables and the power of $\lambda$ by which this contribution is suppressed. Note that in the plots in Sections~\ref{sec:setup}-\ref{sec:LFV} we took the same benchmark value for all the mass insertion, $(\delta_A)_{ij}=0.3$ which is already somewhat suppressed and is of order ${\mathcal O}(\lambda)$. Therefore, the cases where a factor of $\delta^2$ is replaced by $\lambda^2$ should be considered as unsuppressed. 
In cases where there are several dominant contributions in our benchmark studies, we show the one that is the least $\lambda$-suppressed. In cases where there is a strong sensitivity to $\mu$, we show the $\lambda$ scaling both for large and small $\mu$ (of order PeV or TeV, respectively). 

\begin{table}
\begin{center}
\begin{tabular}{ccc}
\hline\hline
\multirow{2}{*}{Observable} & \multicolumn{2}{c}{\qquad\qquad Suppression \qquad\qquad} \\
 &    \multicolumn{2}{c}{large-$\mu$ \qquad \ small-$\mu$}  \\
\hline
\ $u$-EDM (small $t_\beta$) \ &   \  $\delta_{qL}^{13}\delta_{uR}^{13}\to \lambda^6$ \ & \ unaffected \  \\ 
$d$-EDM (large $t_\beta$)& $\delta_{qL}^{13}\delta_{dR}^{13}\to \lambda^4$ & \ \ unaffected \ \  \\ 
$e$-EDM & $\delta_{lL}^{13}\delta_{eR}^{13}\to\lambda^5$ &  \ unaffected \  \\   
\hline
$K$-$\bar K$    &   \multicolumn{2}{c}{$\delta_{qL}^{12}\delta_{dR}^{12}\to \lambda^2$}        \\
$D$-$\bar D$    &    \multicolumn{2}{c}{$\delta_{qL}^{12}\delta_{uR}^{12}\to \lambda^3$}     \\ 
$B$-$\bar B$     &    \multicolumn{2}{c}{$\delta_{qL}^{13}\delta_{dR}^{13}+\ldots \to \lambda^4+\ldots$}    \\
$B_s$-$\bar B_s$     &  \multicolumn{2}{c}{$\delta_{qL}^{23}\delta_{dR}^{23}+\ldots\to \lambda^2+\ldots$}           \\ 
\hline
$\mu\to e \gamma$   & \multicolumn{2}{c}{$|\delta_{lL}^{12}|^2\to 1$}     \\ 
\ $\mu\to e$ conv. \  &  \multicolumn{2}{c}{$|\delta_{lL}^{12}|^2\to  1$}  \\ 
$\mu\to 3e$   & \multicolumn{2}{c}{$|\delta_{lL}^{12}|^2\to 1$}   \\
\hline\hline
\end{tabular}
\end{center}
\caption{\small The effect of a simple horizontal symmetry framework \eqref{eq:squark:masses},
\eqref{eq:slepton:masses}, on observables considered in this work, compared to the benchmark $(\delta_A)_{ij}=0.3\sim{\mathcal O}(\lambda)$ anarchical scenario of Sections~\ref{sec:setup}-\ref{sec:LFV}. For each observable we show which combination of mass insertions controls it and the corresponding $\lambda$ scaling.
 Compared to the benchmark scenario  
 an observable's sensitivity should be considered ``suppressed'' in horizontal symmetry framework only for $\lambda^3$ suppression or higher.
\label{Tab:lambda-supression}}
\end{table}

Though some probes loose significant ground within this framework compared to the anarchical ansatz, others retain their sensitivity. In particular, $K$-$\bar K$ mixing is still robust, and $D$-$\bar D$ mixing is only slightly suppressed compared to our benchmark. Also LFV observables remain sensitive, thanks to the non-hierarchical structure of the $\epsilon_L$ spurions (numerically they are even ${\mathcal O}(1/\lambda^2)$ more sensitive compared to our benchmark scenario in  Sections~\ref{sec:setup}-\ref{sec:LFV}). The strongest loss is suffered by EDM limits in the case of large $\mu$, because these diagrams require a transition from the first to the third generation and back, both in the left and right handed fermion side.
However, we can deduce that many of the probes we have considered are still quite promising.

In summary, having assumed a simple texture for fermion masses, we showed several possibilities for how sfermion masses are affected. Many of the low energy probes remain sensitive, and information from a multitude of channels could begin to distinguish among different flavor models.

\subsection{Radiative Fermion Masses} \label{sec:masses}

Mini-split SUSY and the possibility of large flavor violation at the PeV scale introduces a new opportunity for generating fermion masses. One can imagine a scenario in which only the third generation gets masses at tree level and loops of superpartners induce suppressed masses for the first two generations. In the context of TeV scale SUSY this does not work~\cite{ArkaniHamed:1996zw}, because it requires large flavor violation for squarks and sleptons which in turn generates too large FCNC's. In the context of PeV scale SUSY this opportunity is re-opened, particularly for the first generation~\cite{ArkaniHamed:2012gw}. It is interesting to connect this scenario to low energy probes. This is especially interesting because the mediation of mass, say, from top to up, implies a \emph{lower bound} on required flavor violation. In this section, we first briefly review the 1-loop SUSY corrections to the SM fermion masses and then show that they are directly related to the sizes of (chromo)EDMs. We start by 
focusing on the up quark mass which is very easy to generate in the mini-split SUSY framework. We will then also discuss the possibility of generating the down quark and electron masses, and where in parameter space this attempt pushes us.  

\subsubsection{Radiative Up Mass and the EDM}

\begin{figure}[tb] \centering
\includegraphics[width=0.29\textwidth]{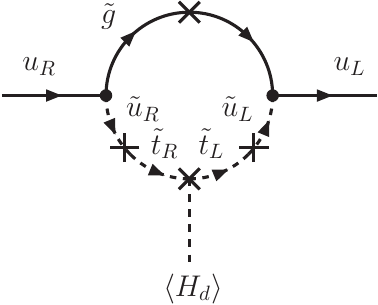} ~~~~~~~ \includegraphics[width=0.29\textwidth]{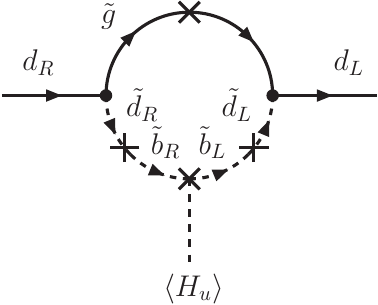} ~~~~~~~ \includegraphics[width=0.29\textwidth]{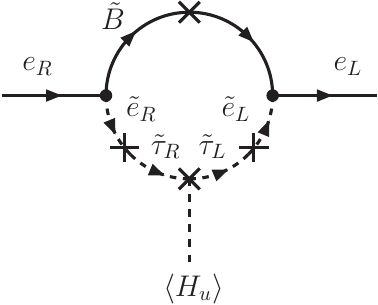}
\caption{\small Flavor violation enhanced 1-loop contributions to the $1^{\rm st}$ generation fermion masses: up quark (left), down quark (middle), and electron (right).}
\label{fig:radiative_mass}
\end{figure}

A generic flavor structure in the sfermion soft masses can lead to large radiative corrections to the SM quark and lepton masses. For instance, the correction to the up quark mass, shown in Fig.~\ref{fig:radiative_mass} (left), is
\begin{equation} \label{eq:mass_up}
\Delta m_u^{\tilde g} = \frac{\alpha_s}{4\pi} \frac{8}{9} \frac{m_{\tilde g} \mu}{m_{\tilde q}^2} ~m_t~ \frac{1}{t_\beta} (\delta^L_{ut} \delta^R_{tu})~.
\end{equation}
For a gluino mass that is a loop factor smaller than the squark mass, $|m_{\tilde g}| \sim 10^{-2}\, m_{\tilde q}$ and a sizable $\mu\sim m_{\tilde q}$, the correction to the up quark mass can be as large as the up quark mass itself, $\Delta m_u^{\tilde g}\sim m_u$,
\begin{equation}
\label{eq:mass_up2}
\Delta m_u \sim 1\mbox{ MeV} \times 
\left(\frac{10^2\, m_{\tilde g}}{m_{\tilde q}}\right)
\left(\frac{\mu}{m_{\tilde q}}\right) 
\left(\frac{1}{\tan\beta}\right)
\left(\frac{\delta^L_{ut} \delta^R_{tu}}{0.3^2}\right)~.
\end{equation}
This means that in mini-split SUSY it is potentially possible to explain the small up quark mass by arguing that it is entirely radiatively generated \cite{ArkaniHamed:1996zw, ArkaniHamed:2012gw}. Numerically, taking a loop factor suppression between squark and gluino masses, $\tan\beta$ can be at most of order a few to get the full up quark mass. Note that an important ingredient is large flavor violation, $(\delta_A)_{ij}\sim {\mathcal O}(1)$,  needed to transmit the enhanced chiral symmetry breaking from the top quark to the up. Hence, in this subsection we will always assume anarchical mass insertions. 

\begin{figure}[tb] \centering
\includegraphics[width=0.96\textwidth]{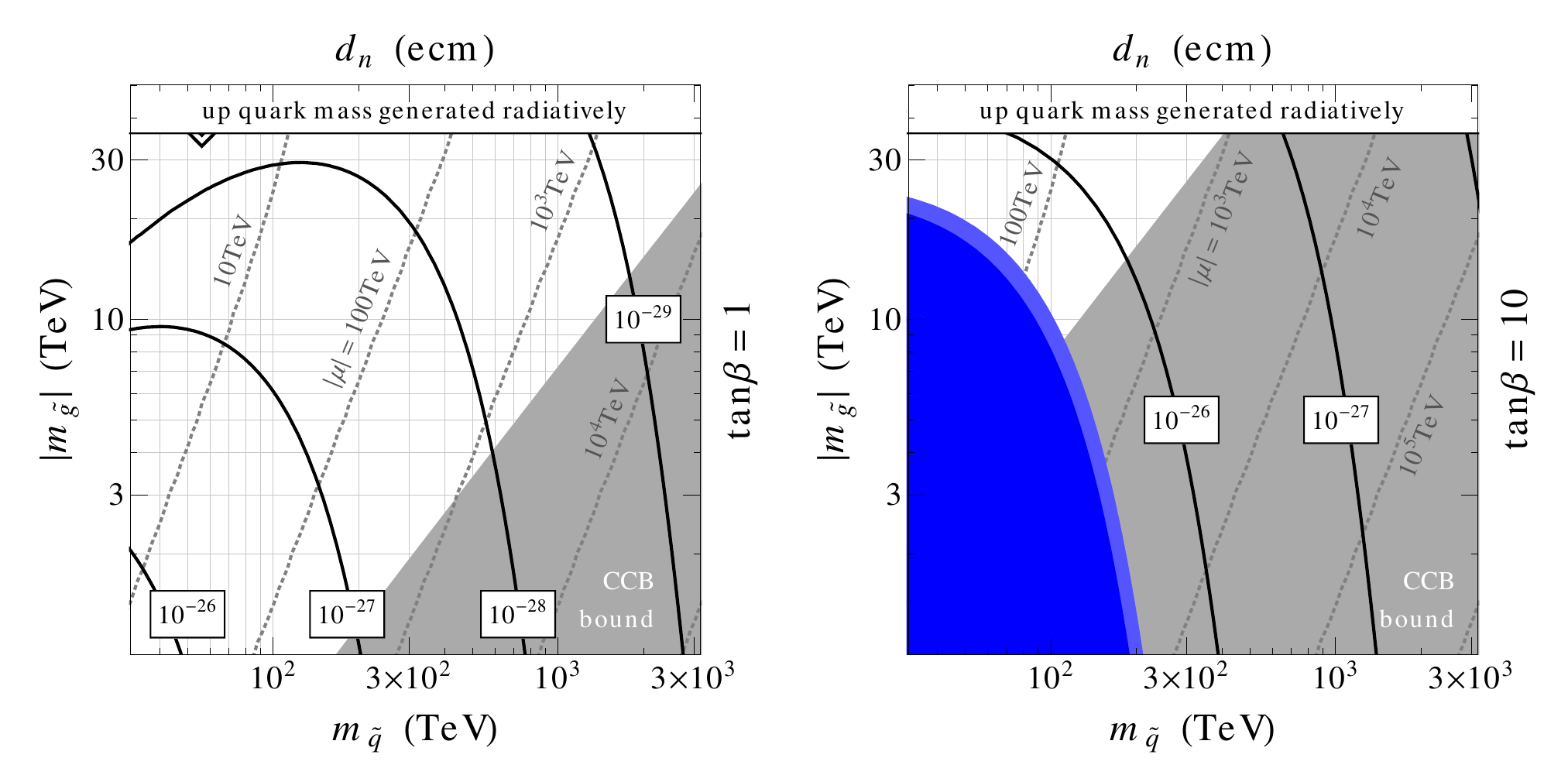}
\caption{\small
Bounds from the neutron EDM in the squark mass $m_{\tilde q}$ vs. gluino mass $|m_{\tilde g}|$ plane in the case of radiative up quark mass generation for $\tan\beta = 1$ (left) and $\tan\beta = 10$ (right). Throughout the plots, the higgsino mass parameter $\mu$ is adjusted such that loop effects can account for the observed up quark mass, as indicated by the gray dotted contours. The dark blue (light blue) shaded regions in the lower left are excluded at the 95\% (90\%) C.L. by current EDM measurements. The gray region in the lower right is excluded by bounds from charge and color breaking.
}
\label{fig:EDMs_rad}
\end{figure}

The diagram in Fig.~\ref{fig:radiative_mass} for the up quark mass is very similar to the leading contribution to the up quark (C)EDM in Fig.~\ref{fig:EDM_diagrams}. The only difference is that an extra photon or gluon gets emitted from the loop. Assuming that the up quark mass comes entirely from SUSY loops naively leads to a tight correlation between the up quark mass and the up quark (C)EDM. However, to make this connection precise one needs to go beyond the diagrams shown in Fig.~\ref{fig:radiative_mass} and Fig.~\ref{fig:EDM_diagrams}, because they have the same phase. To show this consider the limit in which the diagrams in Fig.~\ref{fig:radiative_mass} and Fig.~\ref{fig:EDM_diagrams} are the \emph{only} contributions to the up quark mass and (C)EDMs. Then, if we go to the quark mass eigenbasis in which the up quark mass is real, the dipole diagrams will also be real. This means that the phase $\phi_u$ in Eq. \eqref{eq:EDM_u} is zero and (C)EDMs vanish. As we show next there are, however, other 
contributions beyond  Figs.~\ref{fig:EDM_diagrams},~\ref{fig:radiative_mass} which generate a nonzero phase $\phi_u$ and thus nonzero (C)EDMs.

As an example, consider a model in which the top and charm masses are generated at tree level, and the up mass is generated entirely  from SUSY loops. In this case the first row and column of the quark mass and (C)EDM matrices are zero at tree level. The 11, 12, 13, 21, and 31 entries of the mass matrix are populated at one loop by entries of ${\mathcal O}(m_u)$. The lightest quark mass eigenvalue is then dominated by the 11 entry, and we can neglect the rest of the first row and column. In addition to Figs.~\ref{fig:radiative_mass} and~\ref{fig:EDM_diagrams}, the one-one elements of the up quark mass and (C)EDM matrices receive contributions from ${\mathcal O}(\delta^3)$ mass insertion diagrams. 
Since ${\mathcal O}(\delta^2)$ and ${\mathcal O}(\delta^3)$ diagrams are governed by different loop integrals, this generates nonzero (C)EDMs, with
\begin{equation} \label{eq:EDM_rad}
 \left\{ \frac{d_u(m_{\tilde q})}{e} , \tilde d_u(m_{\tilde q}) \right\} = - \frac{m_u}{m_{\tilde q}^2} ~ 
 \mathrm{Im}\left[ \frac{\delta^L_{uc}\delta^L_{ct}}{\delta^L_{ut}} + \frac{\delta^R_{uc}\delta^R_{ct}}{\delta^R_{ut}}\right]  \times
 ~\left\{ \frac{1}{4} , \frac{231}{64} + \frac{27}{32} \log x \right\}~,
\end{equation}
where $x = |m_{\tilde g}^2|/m_{\tilde q}^2$ (for an early study of connections between (C)EDM and quark mass generation see \cite{Kagan:1994qg}).  For mass insertions $\delta$ not much smaller than 1, the suppression of (C)EDM is not significant.

Note that the loop factor and other factors in \eqref{eq:EDM_u} are now absorbed in the up quark mass $m_u$. The up quark EDM therefore effectively depends only on two parameters, the ratio of mass insertions and on the squark mass. The up quark CEDM is also logarithmically sensitive to the gluino mass. Imposing that $m_u$ is entirely due to SUSY loops reduces the SUSY parameter space by one d.o.f.. Holding all the other parameters fixed, Eq. \eqref{eq:mass_up} can for instance be used to fix the value of $\mu$ from $m_u$. We illustrate the remaining dependence  of neutron EDM, $d_n$, on gluino mass, squark mass and $\tan\beta$ in Fig.~\ref{fig:EDMs_rad}. In it we  set all mass insertions to $|(\delta_A)_{ij}|=0.3$, and also assume generic phases, so that $ \mathrm{Im}[...]=0.6$ in Eq.~(\ref{eq:EDM_rad}).
To calculate $d_n$ we also need the sizes of the down quark (C)EDMs. Here, we are considering a simplified model of radiative quark mass generation with only MSSM fields at the high scale. A relation analogous to Eq. \eqref{eq:EDM_rad} therefore does not apply for the down quark.
The down quark (C)EDM then depends on $\tan\beta$ and $\mu$ explicitly, see Eq.~\eqref{eq:EDM_d}. 

In Fig.~\ref{fig:EDMs_rad} we show two cases, for $\tan\beta=1$ and $\tan\beta=10$, always setting  $|(\delta_A)_{ij}| = 0.3$. The parameter $\mu$ is fixed from \eqref{eq:mass_up}
setting $m_u = 2.5$~MeV, with the resulting values for $|\mu|$ indicated by the dotted gray contours. Expected sensitivities to the neutron EDM of the order of $10^{-28} e$cm can probe this scenario for squark masses near 1000~TeV. 
Note that for large values of $\mu$ a charge and color breaking (CCB) minimum in the MSSM scalar potential can arise where stops acquire a vev. 
In the gray region in the lower right part of the plots denoted by ``CCB bound'', the tunneling rate from our vacuum into the CCB vacuum is faster than the age of the universe~\cite{Kusenko:1996jn}. Note, however, that this bound can be evaded by lowering $\mu$ and increasing the contribution to the up mass, say, by increasing the flavor violating mass insertion beyond 0.3.

As we shall see, the case of $\tan\beta = 10$, can be interesting because it allows to generate simultaneously the up and down quark mass radiatively with comparable mass insertions in the up and down sectors. In this case, the EDM bound is coming dominantly from the down quark CEDM (as opposed to the $\tan\beta=1$ case in which it was the up CEDM which dominated). We discuss this region in parameter space and some of the difficulties that arrise there in greater detail in the next subsection. 

\subsubsection{The Down and Electron Masses}

We will now be more ambitious and try to radiatively generate also the down and electron masses.
As we shall see, 
this will lead us to regions of parameter space that are not typically discussed in the mini-split SUSY framework.  
The corrections to the down quark mass from gluino loop and to the electron masses from bino loop, shown in Fig.~\ref{fig:radiative_mass} (middle) and in Fig.~\ref{fig:radiative_mass} (right) respectively, are given by
\begin{equation}
\begin{split}
\Delta m_d^{\tilde g} &= \frac{\alpha_s}{4\pi} \frac{8}{9} \frac{m_{\tilde g} \mu}{m_{\tilde q}^2} ~m_b~ t_\beta (\delta^L_{db} \delta^R_{bd})~, 
\qquad
\Delta m_e^{\tilde B} = \frac{\alpha_1}{4\pi} \frac{1}{3} \frac{m_{\tilde B} \mu}{m_{\tilde \ell}^2} ~m_\tau~ t_\beta (\delta^L_{e\tau} \delta^R_{\tau e})~,
\end{split}
\end{equation}
where $m_{\tilde B}$ is the bino mass and $m_{\tilde \ell}$ the slepton mass. The correction to the down mass is closely related to that of the up mass 
\begin{equation}
\frac{\Delta m_d}{\Delta m_u} = \frac{m_b}{m_t}\left(\frac{\delta^R_{bd}}{\delta^R_{tu}}\right)t_\beta^2 ~,
\end{equation}
where we have used the fact that left handed mass insertions are equal for up and down. The up to down mass ratio thus sets $\tan\beta\sim 10$ for anarchical mass insertions. From Eq.~(\ref{eq:mass_up2}) we see that for $\tan\beta\simeq10$, getting the overall quark masses correctly requires either that $\mu$ is an order of magnitude above the squark masses, or that the gluino to squark mass ratio is 0.1 rather than $10^{-2}$. Taking a very large $\mu$ is not compatible with the choice of $\tan\beta=10$, unless the Higgs soft masses are also a factor of ten bigger than the squark masses. At the same time, very large values of $\mu$ also lead to dangerous CCB minima in the MSSM scalar potential (see CCB bound in the right plot of Fig.~\ref{fig:EDMs_rad}). On the other hand, splitting the squark and gluino masses by just a single order of magnitude may be a more attractive possibility, since $\tan\beta=10$ requires lighter squark masses for the Higgs mass, as seen in Fig.~\ref{fig:Mh}. However, this spectrum 
is not preferred by anomaly mediated SUSY breaking (AMSB) in its minimal form (adding vector-like matter could help in this respect~\cite{ArkaniHamed:2012gw}). Furthermore, as is also evident from Fig.~\ref{fig:Mh}, such light squarks are generically constrained by meson oscillation limits.

How about the electron mass? The ratio of down to electron masses in this framework is  
\begin{equation}
\frac{\Delta m_d}{\Delta m_e} = 
\frac{8}{3}
\frac{\alpha_s}{\alpha_1}
\frac{m_b}{m_\tau}
\left(\frac{m_{\tilde g}}{m_1}\right)
\left(\frac{m^2_{\tilde l}}{m^2_{\tilde q}}\right)
\left(
\frac{\delta^L_{db}\delta^R_{bd}}{\delta^L_{e\tau}\delta^R_{\tau e}}
\right) ~.
\end{equation}
Taking the AMSB value for the gluino-to-bino ratio and taking similar masses for sleptons and squarks gives a down-to-electron ratio that is about a factor of 30 too big. One is then pushed towards a region where the sleptons are a factor of 5-6 below the squark mass. Given that the squark masses at $\tan\beta\sim10$ need to be around 30-50 TeV, this brings the slepton masses below the 10 TeV scale. As a result this scenario is in strong tension also with $\mu \to e \gamma$. 

We thus conclude that generating the up mass radiatively fits nicely within a flavor anarchic mini-split SUSY framework. This idea will be tested by upcoming searches for hadronic EDMs. 
Even if we are willing to deviate from the AMSB scenario, accommodating the down and electron masses pushes us toward regions with lighter squarks and sleptons which is in conflict with low energy probes such as meson mixing and LFV. Explaining the masses of the first generation radiatively thus requires additional model building.

\subsection{Minimal Flavor Violation}

Minimal flavor violation (MFV) assumes that the Yukawa matrices are the only spurions which break the SM flavor group~\cite{D'Ambrosio:2002ex}. MFV assumption effectively suppresses all flavor violating effects in the quark sector to phenomenologically acceptable levels even for electroweak scale squark masses. The precise value of flavor violation is ambiguous in the lepton sector due to additional mass or Yukawa terms in the neutrino sector~\cite{Cirigliano:2005ck}. For example, the dimension five operator that leads to a Majorana neutrino mass is a spurion that breaks $U(3)_L$ and can have an anarchic flavor structure. Large lepton flavor violation can also arise in the case where the neutrino is Dirac.

We conclude that in mini-split SUSY scenarios which adhere to the MFV ansatz, flavor violating effects are possibly observable in the lepton sector, but not in the quark sector. Note, however, that neutron EDM in the case of small $\mu$   can be mediated by purely flavor conserving interactions (see Sec.~\ref{sec:EDM2loop}).

\section{Discussion}\label{sec:conclusion}

In this work we explored the potential of low energy probes of a mini-split supersymmetry scenario, where sfermions live at around $0.1-1$ PeV. The observables we considered include quark and lepton flavor violating processes as well as electron and neutron electric dipole moments. 
In some cases this leads to notable difference between flavor constraints familiar from studies of natural SUSY. 
For example, in lepton flavor violating processes the penguin contributions are log enhanced. This leads to a deviation from ``dipole-dominance'' in $\mu \to e$ conversion which is commonplace in TeV scale SUSY. The ratio of $\mu\to e\gamma$ and $\mu\to 3e$ still respects the dipole dominance relation in mini-split SUSY. In addition, in mini-split SUSY the limits from meson mixing are essentially independent of the gluino mass and constrain only the squarks, which is obviously not the case for more traditional SUSY spectra. 

Making a flavor anarchic ansatz for squark and slepton mass matrices we found that currently only CP violation in kaon mixing can be sensitive to PeV sfermions. Other probes are not far behind, however, with $D$-$\bar D$ mixing and neutron EDM limits constraining sfermions at 100 TeV. 
Even within a flavor anarchical assumption we cannot predict which flavor and CP violating elements of the sfermion mass matrix will be large and which would be accidentally suppressed. It is thus important to probe the $0.1-1$ PeV scale with as many probes as possible. In this context, the projections for the sensitivity of planned experiments are very exciting. The neutron EDM will become sensitive to squarks at a PeV. The electron EDM and charm mixing will respectively probe sleptons and squarks with masses of around 300 TeV. The upcoming $\mu2e$ experiment  will improve current limits for muon to electron conversion by four orders of magnitude and will probe sleptons at the 100 TeV scale, with further improvements possible with Project X~\cite{Kronfeld:2013uoa}. 

Our results apply beyond the flavor anarchical ansatz for scalar mass matrices. We have for instance also considered the impact of low energy bounds on a a broad variety of models that address the fermion mass hierarchies. For these many low energy precision experiments and rare process searches are still sensitive to the $0.1-1$~PeV scales. As such our results as well as the motivation for low energy experiments become even more robust.

\section*{Acknowledgments} 

\noindent
We would like to thank Alex Kagan, Graham Kribs and Markus Luty for useful discussions.
W.A. thanks Sebastian Ellis for useful correspondence.
Fermilab is operated by the Fermi Research Alliance, LLC under Contract No.
De-AC02-07CH11359 with the United States Department of Energy. J.Z. was supported in part by the U.S. National Science Foundation under CAREER Grant PHY-1151392. We would like to thank KITP for warm hospitality during completion of this work and acknowledge that
this research was supported in part by the National Science Foundation under Grant No. NSF PHY11-25915.
The research of W.A. was supported 
in part by Perimeter Institute for Theoretical Physics. Research at Perimeter Institute is supported by the Government of Canada through Industry Canada and by the Province of Ontario through the Ministry of Economic Development \& Innovation.

\appendix
\section{Log Resummation for Quark CEDMs in Mini-Split SUSY} \label{appendix}

A large hierarchy between gluino and squark masses leads to a large logarithm, $\log(|m_{\tilde g}|^2/m_{\tilde q}^2)$, in the expressions for the quark CEDMs, cf. Eqs. \eqref{eq:EDM_u}, \eqref{eq:EDM_d}. We resum logs of this type by performing a two step matching procedure. In the first step, we integrate out the heavy squarks at the scale $\hat\mu\sim m_{\tilde q}$. This gives us an effective field theory valid for scales $|m_{\tilde g}| \lesssim \hat \mu \lesssim m_{\tilde q}$ where the gluino is still kept as a dynamical degree of freedom. The higher dimensional operators that are generated in the first matching are then evolved down to the low scale $\hat \mu\sim |m_{\tilde g}|$ using renormalization group. Finally, at the low scale,  $\hat \mu \sim |m_{\tilde g}|$, also the gluino is then integrated out from the theory, matching onto the usual CEDM effective Lagrangian \eqref{eq:EDMs_Heff}.

\begin{figure}[tb] \centering
\includegraphics[width=.28\textwidth]{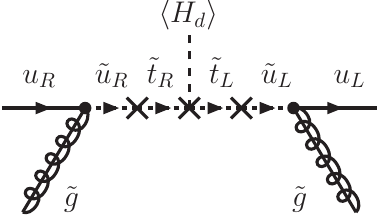} ~~~~~~~~
\raisebox{32pt}{\Large$\Rightarrow$} ~~~~~~~~
\includegraphics[width=.18\textwidth]{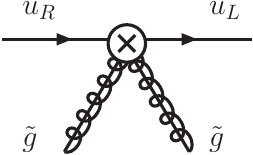}
\caption{\small
An example of a tree level matching that generates the $O_{q\tilde g}, O_{q\tilde g}', O_{q\tilde g}''$  operators (right diagram) when the squarks are integrated out (left diagram) at the scale $\hat \mu\sim m_{\tilde q}$. 
}
\label{fig:app:matching_tree_level}
\end{figure}

We start by integrating out the squarks at the scale $\hat \mu \sim m_{\tilde q}$, which gives an effective theory with SM fields and gauginos. The relevant terms in the effective Lagrangian are
\beq
{\cal L}_{{\rm eff}, \tilde g}=\frac{d_q}{Q_q e}O_q+\tilde d_q \tilde O_q+C_{q\tilde g} O_{q\tilde g}+C_{q\tilde g}' O_{q\tilde g}'+C_{q\tilde g}'' O_{q\tilde g}''.\label{Leff_tildeg}
\eeq
The first two terms are the usual dimension 5 (C)EDM operators 
\begin{eqnarray} \label{eq:op_EDM}
 O_q &=& -\frac{i}{2} eQ_q ~(\bar q_\alpha \sigma^{\mu \nu} \gamma_5 q_\beta) ~F_{\mu\nu} ~ \delta_{\alpha\beta}~, \\
 \tilde O_q &=& -\frac{i}{2} g_s ~(\bar q_\alpha \sigma^{\mu \nu} \gamma_5 q_\beta) ~G_{\mu\nu}^a ~ T_{\alpha\beta}^a ~.
\end{eqnarray}
The remaining three operators in \eqref{Leff_tildeg} are of dimension 6 and contain two quarks and two gluino fields. In  \eqref{Leff_tildeg} we only keep the CP violating operators, since they are the ones that are potentially important for the running of the CEDM operators. The notation that we use is
\begin{eqnarray}
O_{q\tilde g} &=& g_s^2 \frac{1}{m_{\tilde g}} \Big[ (\bar q_\alpha \tilde g_a)(\bar{\tilde g}_b \gamma_5 q_\beta) + (\bar q_\alpha \gamma_5 \tilde g_a)(\bar{\tilde g}_b q_\beta) \Big] ~f_{abc} T_{\alpha\beta}^c~, \\
O_{q\tilde g}^\prime &=& i g_s^2 \frac{1}{m_{\tilde g}} \Big[ (\bar q_\alpha \tilde g_a)(\bar{\tilde g}_b \gamma_5 q_\beta) + (\bar q_\alpha \gamma_5 \tilde g_a)(\bar{\tilde g}_b q_\beta) \Big] ~d_{abc} T_{\alpha\beta}^c~, \\
O_{q\tilde g}^{\prime\prime} &=& i g_s^2 \frac{1}{m_{\tilde g}} \Big[ (\bar q_\alpha \tilde g_a)(\bar{\tilde g}_b \gamma_5 q_\beta) + (\bar q_\alpha \gamma_5 \tilde g_a)(\bar{\tilde g}_b q_\beta) \Big] ~\delta_{ab} \delta_{\alpha\beta}~, \label{eq:op_Opp}
\end{eqnarray}
where $\alpha, \beta = 1 \dots 3$ and $a,b,c = 1 \dots 8$ are color indices and $f_{abc}$ ($d_{abc}$) are the totally anti-symmetric (symmetric) $SU(3)$ structure constants. The normalization factor of $g_s^2 / m_{\tilde g}$ is introduced for later convenience.

\begin{figure}[tb] \centering
\includegraphics[width=.25\textwidth]{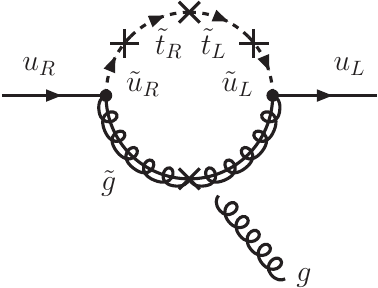} ~~~~~~~
\raisebox{56pt}{\Large$\Rightarrow$}~~~~~~~
\raisebox{12pt}{\includegraphics[width=.25\textwidth]{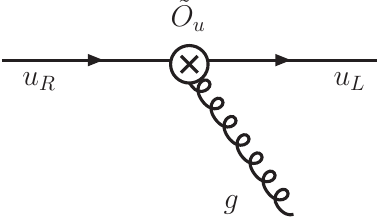}}~~
\raisebox{56pt}{\Large$+$}~~
\includegraphics[width=.25\textwidth]{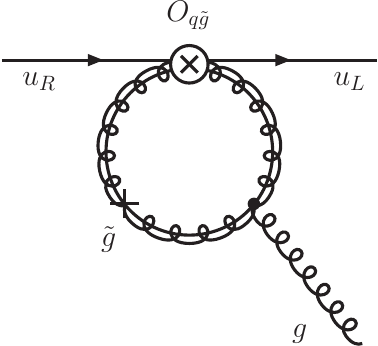}
\caption{\small
An example of 1-loop level matching that generates the $\tilde O_u$ operator at the scale $\hat \mu\sim m_{\tilde q}$. 
}
\label{fig:app:matching_loop_level}
\end{figure}

The Wilson coefficients of the quark-gluino operators at the high scale $\hat \mu\simeq m_{\tilde q}$ are obtained by integrating out the squarks in the tree level diagrams such as shown in Fig. \ref{fig:app:matching_tree_level}, giving  
\begin{eqnarray}
\Big\{ C_{u\tilde g} , C_{u\tilde g}^\prime, C_{u\tilde g}^{\prime\prime} \Big\} &=&  \frac{|\mu m_{\tilde g}|}{m_{\tilde q}^4} m_t \frac{1}{t_\beta} ~|\delta_{ut}^L \delta_{tu}^R|~ \sin\phi_u ~\left\{ -\frac{1}{2} , \frac{1}{2}, \frac{1}{6} \right\} ~,\\
\Big\{ C_{d\tilde g} , C_{d\tilde g}^\prime, C_{d\tilde g}^{\prime\prime} \Big\} &=&  \frac{|\mu m_{\tilde g}|}{m_{\tilde q}^4} m_b ~t_\beta ~|\delta_{db}^L \delta_{bd}^R|~ \sin\phi_d ~\left\{ -\frac{1}{2} , \frac{1}{2}, \frac{1}{6} \right\} ~.
\end{eqnarray}
The (C)EDM Wilson coefficients at $\hat \mu\simeq m_{\tilde q}$ are then obtained by performing the matching at 1-loop with representative diagrams shown in Fig. \ref{fig:app:matching_loop_level},  giving for the 
case of the up quark
\begin{eqnarray}
 \left\{ \frac{3d_u}{2e} , \tilde d_u \right\} &=& \frac{\alpha_s}{4\pi} \frac{|\mu m_{\tilde g}|}{m_{\tilde q}^4} m_t \frac{1}{t_\beta} ~|\delta_{ut}^L \delta_{tu}^R|~ \sin\phi_u ~\left\{ -\frac{4}{3} , -\frac{59}{6} \right\} ~, 
\end{eqnarray}
while for the case of down quark one obtains
\begin{eqnarray}
 \left\{ -\frac{3d_d}{e} , \tilde d_d \right\} &=& \frac{\alpha_s}{4\pi} \frac{|\mu m_{\tilde g}|}{m_{\tilde q}^4} m_b ~t_\beta ~|\delta_{db}^L \delta_{bd}^R|~ \sin\phi_d ~\left\{ -\frac{4}{3} , -\frac{59}{6} \right\} ~.
\end{eqnarray}
Note that no large logs appear in these Wilson coefficients.

\begin{figure}[tb] \centering
\includegraphics[width=.26\textwidth]{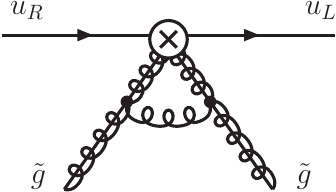} ~~~~~~
\includegraphics[width=.26\textwidth]{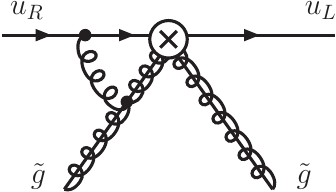} ~~~~~~
\includegraphics[width=.26\textwidth]{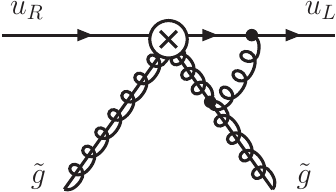} \\[16pt]
\includegraphics[width=.26\textwidth]{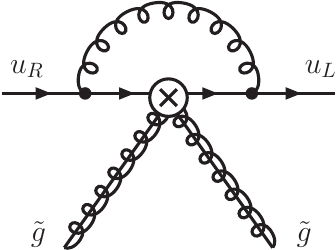} ~~~~~~
\includegraphics[width=.26\textwidth]{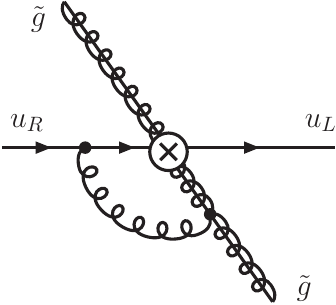} ~~~~~~
\includegraphics[width=.26\textwidth]{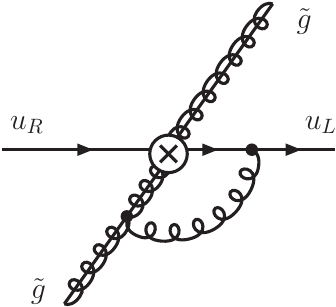}
\caption{\small
The diagrams that determine the running and mixing of the operators $O_{q\tilde g}$, $O_{q\tilde g}^\prime$, and $O_{q\tilde g}^{\prime\prime}$ at the 1-loop level.
}
\label{fig:NLO}
\end{figure}

The operators in (\ref{eq:op_EDM})--(\ref{eq:op_Opp}) mix under renormalization. In particular $O_{q\tilde g}$ mixes at 1-loop into the CEDM operator $\tilde O_q$ with the relevant diagram shown in Fig.~\ref{fig:app:matching_loop_level} (right-most diagram). The other four fermion operators, $O_{q\tilde g}^\prime$ and $O_{q\tilde g}^{\prime\prime}$, on the other hand, cannot mix at 1-loop into $\tilde O_q$ due to their symmetric color structure. The running and mixing of the four fermion operators is determined by the diagrams shown in Fig.~\ref{fig:NLO}. For the 1-loop anomalous dimension matrix $\hat \gamma$ that determines the running and mixing of the dipole operators $O_q$ and $\tilde O_q$ with the operators $O_{q\tilde g}$, $O_{q\tilde g}^\prime$, and $O_{q\tilde g}^{\prime\prime}$, we find
\begin{equation}
 \hat \gamma = \frac{\alpha_s}{4\pi} \hat \gamma_0 = \frac{\alpha_s}{4\pi} \begin{pmatrix} 8/3 & 0 & 0 & 0 & 0 \\ 32/3 & 4/3 & 0 & 0 & 0 \\ 0 & 12 & -52/3 & 18 & 12 \\ 0 & 0 & 10/3 & -16 & 0 \\ 0 & 0 & 8 & 0 & -34 \end{pmatrix}~.
\end{equation}
%
The results recently presented in~\cite{Fuyuto:2013gla} are consistent with this anomalous dimension matrix.\footnote{We thank the authors of~\cite{Fuyuto:2013gla} for pointing out an error in the first version or this work.}

The Wilson coefficients then evolve according to
\begin{eqnarray} \label{eq:running}
 && \left( \frac{d_q(|m_{\tilde g}|)}{e Q_q},\tilde d_q(|m_{\tilde g}|), C_{q \tilde g}(|m_{\tilde g}|), C_{q \tilde g}^\prime(|m_{\tilde g}|), C_{q \tilde g}^{\prime\prime}(|m_{\tilde g}|) \right)^{\rm T} \nonumber \\
 && = \eta^{\frac{\hat \gamma_0^{\rm T}}{2 \beta_0}} \left( \frac{d_q(m_{\tilde q})}{e Q_q},\tilde d_q(m_{\tilde q}), C_{q \tilde g}(m_{\tilde q}), C_{q \tilde g}^\prime(m_{\tilde q}), C_{q \tilde g}^{\prime\prime}(m_{\tilde q}) \right)^{\rm T}  ~,
\end{eqnarray}
with $\eta$ given by
\begin{equation}
 \eta = \frac{\alpha_s(m_{\tilde q})}{\alpha_s(|m_{\tilde g}|)} = \frac{1}{1 - \beta_0 \frac{\alpha_s(|m_{\tilde g}|)}{4\pi} \log(|m_{\tilde g}|^2/m_{\tilde q}^2)} ~.
\end{equation}
In the above expressions, $\beta_0$ is the 1-loop coefficient of the QCD beta function. Taking into account that there is still a contribution of a gluino to $\alpha_s$ running, one has~\cite{Jones:1974pg}
\begin{equation} \label{eq:beta_s}
 \beta_s = - \frac{\alpha_s}{4\pi} \beta_0 = - \frac{\alpha_s}{4\pi} (11 - \frac{2}{3} n_f - 2 n_{\tilde g}),
\end{equation}
where $n_f$ and $n_{\tilde g}$ are the numbers of active quarks and gluinos, i.e. $n_{\tilde g} = 1(0)$ above (below) the gluino threshold. 

Explicitly, for the quark CEDMs at the gluino scale, $\hat \mu\simeq |m_{\tilde g}|$, we get from~(\ref{eq:running})  
\begin{equation} \label{eq:CEDM_resummed}
 \tilde d_q(|m_{\tilde g}|) = \eta^\frac{2}{15} \tilde d_q(m_{\tilde q}) + C_{q\tilde g}(m_{\tilde q}) \sum_{i=1}^4 h_i \eta^{a_i} + C_{q\tilde g}^\prime(m_{\tilde q}) \sum_{i=1}^4 h_i^\prime \eta^{a_i} + C_{q\tilde g}^{\prime\prime}(m_{\tilde q}) \sum_{i=1}^4 h_i^{\prime\prime} \eta^{a_i} ~,
\end{equation}
with the ``magic numbers''
\begin{equation}
\begin{array}{llll}
 a_1 = 2/15 ~, & a_2 = -0.706 ~, & a_3 = -2.125~,, & a_4 = -3.903~, \\
 h_1 = 0.961 ~, & h_2 = -0.759 ~, & h_3 = -0.141~, & h_4 = -0.061~, \\
 h^\prime_1 = 0.185~, & h^\prime_2 = -0.283 ~, & h^\prime_3 = 0.090~, & h^\prime_4 = 0.009~, \\
 h^{\prime\prime}_1 = 0.218~, & h^{\prime\prime}_2 = -0.226 ~, & h^{\prime\prime}_3 = -0.088 ~, & h^{\prime\prime}_4 = 0.096~. \\ 
\end{array}
\end{equation}
Since we are working at leading log approximation, integrating out the gluino at the scale $\hat \mu\simeq |m_{\tilde g}|$ does not give further threshold corrections to the CEDM. In~(\ref{eq:CEDM_resummed}) the large log is resummed by the renormalization group evolution.
Expanding this result to leading order in $\alpha_s$, we recover the fixed order results in~(\ref{eq:EDM_u}) and~(\ref{eq:EDM_d}). Numerically, the correction from the RG running can be significant. For instance, for $|m_{\tilde g}|=3$~TeV and $m_{\tilde q}\simeq 10^3$~TeV it is a  $\sim85\%$ correction to the unresummed result.

Note that all quantities entering the initial conditions of the Wilson coefficients at the squark scale, $\hat\mu=m_{\tilde q}$, also need  to be evaluated at the same scale $m_{\tilde q}$. This is in particular true for $\alpha_s$, the top and bottom masses and also the gluino mass. The 1-loop running of $\alpha_s$ in the presence of a dynamical gluino is in~(\ref{eq:beta_s}). The anomalous dimension of the quark masses does not get altered at 1-loop by the presence of a gluino, and thus
\begin{equation}
 \frac{m_q(m_{\tilde q})}{m_q(|m_{\tilde g}|)} = \eta^\frac{\gamma_{m_q}}{2 \beta_0} = \eta^\frac{4}{5} ~. 
\end{equation}
For the 1-loop running of the gluino mass we find   
\begin{equation}
 \frac{m_{\tilde g}(m_{\tilde q})}{m_{\tilde g}(|m_{\tilde g}|)} = \eta^\frac{\gamma_{m_{\tilde g}}}{2 \beta_0} = \eta^\frac{9}{5} ~. 
\end{equation}
In all the plots we show, the gluino mass refers to $m_{\tilde g}(|m_{\tilde g}|)$. 
Below the gluino threshold, $\hat\mu \lesssim |m_{\tilde g}|$, the evolution of $\alpha_s$, $m_q$, as well as the quark (C)EDMs is governed by the respective SM RGEs.

\section{Loop Functions} \label{app:loopfunctions}

\noindent
The loop functions that enter the wino contributions to $\mu \to e \gamma$ read
\begin{eqnarray}
 g_1(x) &=& \frac{1+16x+7x^2}{(1-x)^4} + \frac{2x(4+7x+x^2)}{(1-x)^5} \log x~,\\[16pt]
 g_2(x,y) &=& - \frac{11 + 7(x+y) -54xy +11(x^2y+y^2x) + 7x^2y^2}{4(1-x)^3(1-y)^3} \nonumber \\
 && + \frac{x(2+6x+x^2)}{2(1-x)^4(y-x)} \log x + \frac{y(2+6y+y^2)}{2(1-y)^4(x-y)} \log y ~,\\[16pt]
 g_3(x,y) &=& - \frac{40 -33(x+y) +11(x^2+y^2) + 7(x^2y+y^2x) - 10xy}{4(1-x)^3(1-y)^3} \nonumber \\
 && + \frac{2+6x+x^2}{2(1-x)^4(y-x)} \log x + \frac{2+6y+y^2}{2(1-y)^4(x-y)} \log y ~.
\end{eqnarray}

\noindent
The loop functions that enter the $Z$ penguin contributions to $\mu \to e$ conversion read
\begin{eqnarray}
 f_1(x,y) &=& \frac{x^3(3-9y)+(y-3)y^2+x^2(3y-1)(1+4y)+xy(y(13-11y)-4)}{2(1-x)^2(1-y)^2(x-y)^2} \nonumber \\
 && + \frac{x(2x^3+2y^2+3xy(1+y)-x^2(1+9y))}{(1-x)^3(x-y)^3} \log x \nonumber \\
 && + \frac{y^2(y+x(7y-5)-3x^2)}{(1-y)^3(x-y)^3} \log y ~,\\[16pt] 
 f_2(x,y) &=& \frac{x^3(1-3y)+3(y-3)y^2-x(y-3)y(y+4)+x^2(y(13-4y)-11)}{2(1-x)^2(1-y)^2(x-y)^2} \nonumber \\
 && + \frac{x(2x^3+2y^2+3x^2(1+y)-xy(9+y))}{(1-x)^3(y-x)^3} \log x \nonumber \\
 && + \frac{y^2(x^2+x(7-5y)-3y)}{(1-y)^3(y-x)^3} \log y ~,\\[16pt] 
 f_3(x,y) &=& -\frac{12(x+y+x^2+y^2+x^2y+y^2x-6xy)}{(1-x)^2(1-y)^2(x-y)^2} \nonumber \\
 && + \frac{24x(x^2-y)}{(1-x)^3(y-x)^3} \log x + \frac{24y(y^2-x)}{(1-y)^3(x-y)^3} \log y ~. 
\end{eqnarray}



\end{document}